\documentclass[11pt,a4paper]{article}
\usepackage{t1enc}
\usepackage[latin1]{inputenc}
\usepackage[english]{babel}
\usepackage{harvard}
\usepackage{graphicx}
\newcommand{\Z}{\mbox{\sf Z\hspace{-0.8ex}Z}}

\newcommand{\R}{\mbox{\rm I\hspace{-0.33ex}R}}

\newcommand{\beq}{\begin{equation}}
\newcommand{\eeq}{\end{equation}}
\newcommand{\beqarr}{\begin{eqnarray}}
\newcommand{\eeqarr}{\end{eqnarray}}
\newcommand{\beqa}{\begin{eqnarray*}}
\newcommand{\eeqa}{\end{eqnarray*}}
\newtheorem{principle}{Principle}

\begin{document}
\thispagestyle{empty}

\title{On the geometry of cosmological model building}
\author{Erhard Scholz\footnote{University Wuppertal, Faculty of Mathematics and Natural Sciences,
  scholz@math.uni-wuppertal.de}}
\date{November 16, 2005 }
\maketitle
%\vspace{2.5mm}

\begin{abstract}
This article analyzes the present anomalies of  cosmology  from the point of view  of integrable Weyl geometry. It uses P.A.M. Dirac's  proposal for  a weak extension of general relativity, with some small adaptations.  Simple models  with interesting geometrical and physical properties, not belonging to the
Friedmann-Lema\^{\i}tre class, are studied  in this frame. Those with positive spatial curvature 
(Einstein-Weyl universes)  go well together with observed mass density $\Omega _m$,  CMB, supernovae Ia data,  and  quasar frequencies. They  suggest a physical role for an equilibrium state of the Maxwell field proposed by I.E. Segal in the 1980s (Segal background) and for a  time invariant balancing condition of vacuum energy density. The latter leads to a surprising agreement with the  BF-theoretical calculation proposed by  C. Castro \cite{Castro:ADSetc}. 
\end{abstract}

\subsection*{1. Introduction}
The current standard model of  cosmology is characterized by  both, striking successes and grave problems. Most impressive  recent  successes consist of a comparably sharp  determination of the two basic model parameters
 $(\Omega _m, \Omega _{\Lambda }) \approx (0.25, 0.75)$  by supernovae data,  presumed consistency with the estimation of matter density  and with the observed anisotropies of the cosmic microwave background 
(CMB).\footnote{See, among a flood of literature,  \cite{Groen:Universe,Carroll:Constant,Peebles:Constant}.} 
More settled  achievements of the model are the explanation of cosmological redshift by an expansion of the space sections of  space-time, the characterization of the  CMB as a remnant of a hot early stage of the universe, and the calculation of  ratios of the light elements in the cosmos derived from equilibria  in the hypothetical primordial synthesis of atomic nuclei.
All this was  instrumental for a broad acceptance of the expansionary cosmological models in the 1960s and is now  part of the recent history of physics \cite{Kragh:Cosmos}.

On the other hand, the relative mass density $\Omega _m$ is an order of magnitude larger than the relative density of baryonic matter, $\Omega _b \approx 0.02 $, 
maximally  allowed by the theory of primordial nucleosynthesis. Experimental and theoretical search for new kinds of ``exotic'' dark matter has so far been without success,  while the expectation  that new insights of particle physics may some day solve the problem   is still widely held \cite{Peebles:Open,Peebles:Dark,Ellis:Future}. 
Recent  studies  indicate, however,  that exotic dark matter of any kind in the high amount demanded by the present mass density data would contradict other astronomical evidence \cite{Overduin/Wesson:DarkMatter}.  In this sense, the postulate of {\em exotic} dark matter has turned into an {\em   anomaly} of  the  {\em  standard model of cosmology} (SMC) .

 The inconsistency does not directly arise   from the empirical data. Astronomers know well  that molecular hydrogen is difficult to trace observationally  by absorption or emission. Dynamically determined mass density on larger scales is constrained at the moment by $0.1 \leq \Omega _m \leq 1$   \cite[26ff.]{Carroll:Constant}. It can be restricted by comparison with estimations by  other methods   to  $0.15 \leq \Omega _m \leq 0.3$ \cite{Peebles:Probing}. The wider interval, or an even moderately larger one, seems to be consistent with the  bounds for intergalactic molecular hydrogen as far as it  can be estimated from absorption  data  \cite{Wszolek:Limits}.  Inconsistency  arises only from  assuming the  {\em hypothesis of primordial nucleosynthesis}. Theories of structure formation in expanding universes also build upon  the existence  of non-baryonic dark matter as a consequences of this hypothesis. But they  remain  highly problematic in  themselves. The exotic dark matter anomaly gives  thus testimony of  broad {\em  empirical evidence against} the hypothesis of primordial nucleosynthesis.  Although the last word in this question has not been spoken, we shall not make use of it.

Some experts of the field consider the paradoxical features of a  physically realistic interpretation of $ \Omega _{\Lambda } \approx 0.7$, the  other parameter of the SMC, as even more problematic 
\cite{Carroll:Constant,Straumann:Myst,Giulini/Straumann}.  $\Omega _{\Lambda }$  is  interpreted as a contribution of  a {\em dynamical dark energy}  to the  right hand side of the Einstein equation, due to fluctuations of the quantum vacuum or, alternatively, to an unseen stuff (``quintessence'') mathematically modelled by a scalar field. The second version has the  `advantage' that a time dependence can be built into the model by a free choice of a (time-dependent) potential. In any case, the actual parameters of the standard model indicate a monotonous increase of $ \Omega _{\Lambda } (t)$ over the cosmic time parameter $t$, while  $\Omega _{\Lambda } +  \Omega _m$ remains constant   equal $1$, just right to assure flatness of the model's space sections.  Two implications of the time dependence of $ \Omega _{\Lambda } (t)$ are nevertheless particularly irritating. Firstly it indicates an {\em accelerated} expansion of spatial sections with the consequence of a monotonous reduction of the causally accessible parts of the universe in the progression of cosmic time. Although this is no problem for SMC as a purely mathematical model, it  has unpleasing philosophical consequences if one insists on a realistic interpretation of the model. Our macroscosmos seems to fragmentize into more and more unconnected
 parts.\footnote{One is tempted to read this property as an unintendedly ironic counterimage of evolutionary trends of  late modern  society. }  Secondly,  the  ratio $\Omega _{\Lambda } : \Omega _{m } \approx 0.7 : 0.3$ is crucial for supplying at least roughly realistic conditions of structure formation in an expanding cosmos, but  holds only  close  to the present time. That leads to the {\em cosmic coincidence} problem much discussed in the literature, bringing different kinds of ``anthropic principles'' into the game. The attempts to explain the dynamical dark energy   by electrodynamic (or more general) quantum fluctuations lead to an  error of 120 orders of magnitude, apparently  the ``worst prediction in the history of
 physics''.\footnote{Giulini and Straumann characterize a toy  estimation of W. Pauli by this qualification, although its error is ``only''  27 orders of magnitude \cite[9]{Giulini/Straumann}.}

Other problems abound. 
Astronomical observations show {\em  a lack of increase of   mean metallicity of  galaxies} over cosmological time 
\cite{Corbin:QSOs,Grebel:metal_poor,Pagel:Nucleo}, contrary to what one would expect from a globally evolving cosmos. Not much recognized\footnote{See, e.g., the otherwise beautiful presentation in \cite[363]{Carroll:Spacetime}.}
 but at least as irritating for the standard picture is  a result of  lattice gauge theory, demonstrating the inconsistency of   the {\em hypothesis of the electroweak phase transition} in the early universe with the present standard model of elementary particle physics \cite{Fodor:ew_phase}.
 This result also feeds doubts  with respect  to   the much ``earlier'' {\em inflationary  epoch} of the universe.
 Other authors consider the {\em  impossibility} to come to satisfying models of {\em structure formation} in an expanding universe as an indicator for a crisis of the standard aproach 
\cite{Ostriker:Darkmatter}. Finally, correlation studies have   shown that the {\em anisotropies of the CMB} can no longer be considered  to result from primordial effects only. At least a part of them is due to foreground inhomogeneities in the course of the radiation through galaxies and clusters (Sunyaev-Zeldovich effect) \cite{Shanks:WMAP}. At the moment, the questions   is about how much of the anisotropies are due to foreground effects, but it may turn out that the  whole anisotropy signal is due to such ``foreground'' effects. 
 
These are important questions  and may become even more so in the future.  We  agree with P. Steinhardt, J. Ostriker, N. Turok and others in considering the great advances of observational cosmology of the last decades to give new incentives for revisiting  the foundations of    cosmological model building, rather than feeling assured of the impeccability  of the standard frame \cite{Steinhardt:Cyclic,Ostriker:Darkmatter}.  

 Different to Steinhardt and Turok, who  propose a   cylical extension of  standard cosmology based on quantum physical hypotheses, we choose here to reconsider the foundations of relativistic  geometry from a  more conceptual  point of view which deals mainly with 
semi-classical geometrical considerations. We use H. Weyl's old idea of a scale gauge geometry for extending   general relativity and investigate  consequences for the construction of cosmological space-time. Our extension will be much weaker than Weyl's original proposal. It uses   {\em integrable Weyl geometry} (IWG) only, similar to Dirac's attempts of the 1970s. It will turn out that already this  tiny theoretical modification of Riemannian geometry is well adapted to the task of analyzing the present problems in cosmology. Among others it  allows to form very simple cosmological models which  shed light on the present constellation of anomalies.

 In the Weyl geometrical approach cosmological redshift can be mathematically characterized by a Weylian length (scale) connection, here called {\em Hubble connection}. Then space expansion becomes mathematically equivalent to a gauge effect which {\em may be} physical, but {\em need not be} so. If it is not,  the cosmic microwave background has to be explained by another cause than  at present. For positive spatial curvature  a natural alternative was proposed by I.E. Segal in the 1980s 
by an equilibrium background state of the quantized Maxwell field  \cite{Segal:CMB1}.  In this case, anisotropies would arise exclusively from inhomogeneities in the ``foreground''. 

Our modification of  the conceptual framework of standard cosmology is minimal and the deviation of general relativity on the small scale (solar system or galactic) is negligible; but it implies considerable changes of geometry and physics in larger regions. Most importantly, we can  form space-times with cosmological redshift but  without big bang. It depends on the gauge perspective, whether there arises an initial singularity or not. Present data of supernovae Ia, mass density, cosmic microwave background, and quasar frequencies are consistent with a   Weyl geometric version of  the Einstein universe. This approach leads to  a time homogeneous cosmological geometry in large means, where evolution becomes a local, or better a regional feature  in parts of the cosmos only. 

In the following section we start with a reminder of   a conservative extension of GRT, similar to the one proposed by P.A.M. Dirac  in the 1970s (section 2). A broader introduction to the background of integrable Weyl geometry  used in this paper can be found in appendix I.   We  look at Robertson-Walker geometries from our vantage point and  characterize cosmological redshift by a Weylian length connection (section 3). Particularly simple integrable Weyl geometries useful for cosmological model building ({\em Weyl universes}) are introduced und studied (section 4).  
 In our approach the data lead to  positively curved spatial sections ({\em Einstein-Weyl models}).  That suggests to reconsider Segal's alternative explanation of the CMB. The metrical parameter of Einstein-Weyl universes (one only)  is sharply constrained by present mass density values (section 5). The resulting model predicts the redshift-luminosity data data of supernovae SNI$_a$
as well as the model class of the standard approach. Anisotropies with a peak about the momentum $l \approx 200$ are results of ``foreground'' inhomogeneities of the gravitational field around galaxy clusters and superclusters. Moreover, Einstein-Weyl models  give a surprising geometrical view of quasar frequencies (section 6).   All in all, the new model class sheds light on the strategic decisions of theory construction in the standard approach   (section 7).

%\newpage
\subsection*{2. A conservative extension of general relativity }
%%%%%%%%%%%%%%%%%%
{\bf   Dirac's extended IWG } \\ \noindent
We follow P.A.M. Dirac's  proposal for using integrable Weyl geometry in general relativity  \cite{Dirac:1973}.   After choice of coordinates $x=(x^0, x^1,
x^2, x^3)$,  a {\em  Weylian  metric} is locally expressed by a semi-Riemannian metric $g$ and a differential 1-form $\varphi$. $g$ will be  called the {\em Riemannian component} of the 
metric,
\[  g  = (g_{ij}), \; \; \, ds^2 = \sum_0^4 g_{ij} dx^i dx^j \; . \]
It expresses metrical relations  with respect to a chosen {\em gauge} and allows to compare metrical quantities directly only if they are measured at the same 
``point'' (event). The additional  1-form,  
\[ \varphi  = (\varphi _i) , \; \; \; \, \varphi = \sum \varphi_i dx^ i  \; , \]
encodes information of how to compare metrical quantities between different points of the manifold. It is called the  {\em length} or {\em scale connection} of the Weylian metric.  The pair $(g, \varphi)$ defines  a
{\em gauge} of the metric. It can be changed to another one, $(\tilde{g}, \tilde{\varphi}) $,  by a  {\em gauge transformation}  given by 
\beq \tilde{g}(x) = \Omega ^2(x) \, g(x) = e^{2 \Phi (x)} g(x) \, , \;\; \tilde{\varphi} = \varphi - d  \log \Omega =
\varphi - d \Phi \, ,
 \eeq
with a real valued function $\Phi$ on a local neighbourhood. That means, the Riemannian component is rescaled  and the
length connection modified by subtracting  the differential of  $\log \Omega $
of the rescaling function $\Omega (x) = e^{\Phi (x)}$. The gauge transformation is defined in such a way  that,  by means of  integrals of the scale connection,  metrical quantities at different points may be compared independently of gauge (appendix I, equs. (\ref{length-transfer}) to ( \ref{calibration-transfer})).

We generally assume the integrability of the scale connection, $d \varphi = 0$,  by reasons given below; that is we work in {\em integrable Weyl geometry} (IWG).  In this case, a {\em semi-Riemannian gauge} with vanishing scale connection, $(\tilde{g}, 0)$, exists. In the physics literature it is often called {\em Einstein gauge}. For more details on the geometrical properties of Weyl geometry see appendix I. Please note the  warning with respect to the {\em different sign convention} for the gauge transformation of the differential form used in most of the physical literature on IWG.
%%%%%%%%%%%%%%%%
\\[0.5ex] \noindent
{\bf Free fall trajectories}
\\ \noindent
Drawing upon Dirac's gauge scale covariant geodesics (cf. app. I), we can directly take over the basic principle  of physical geometry used in GRT:
\begin{principle} \label{trajectories}
The trajectories of freely falling particles are described by timelike, scale gauge covariant geodesics in IWG. The motion of photons is described by gauge covariant null-geodesics.
\end{principle} 
The  parametrization of geodesics  demanded by the postulate is essential. Transfer of mass and of photon energy (with respect to observer fields) is characterized by  {\em gauge covariant}  geodesics. This is seen differently in the literature which  applies Weyl geometry but  does not follow the Dirac tradition. In the next paragraph it becomes clear why Dirac's approach is preferrable for physically meaningful usage of integrable Weyl geometry.
%%%%%%%%%%%%%%%%
\\[0.5ex] \noindent
{\bf Compatibility with quantum mechanics}
\\ \noindent
Principle \ref{trajectories} is close to two axioms in the constructive axiomatics of Ehlers, Pirani and Schild 
\cite{Ehlers/Pirani/Schild}. This ground breaking approach worked with Weyl's {\em gauge invariant} geodesics. Dirac's modification \cite{Dirac:1973} was not yet published. The literature following Ehlers e.a. continued in this tradition,  cf. \cite{Audretsch_ea,Perlick:Observerfields}.  That led to unnecessary obstacles for the use of IWG in general relativity.  In particular, 
 the compatibility condition  of quantum mechanics and Weyl geometry derived in \cite{Audretsch_ea} seemed to suggest  that   Riemann gauge must be chosen and that the Weyl geometric extension is redundant. 

Audretsch/G\"ahler/Straumann studied Klein/Gordon matter fields $ \psi$ on a Weylian manifold (with vanishing topological obstructions) and assumed them to  evolve, roughly speaking, along geodesic paths. More precisely  they
  considered the 
 WKB development of $\psi$,  the series development of the solutions in rising powers of $\hbar$, 
\[   \psi = e^{\frac{i S}{\hbar}} \left(  \psi_0 + \frac{\hbar}{i} \psi_1 + \ldots  \right) \; . \] 
Their compatibility criterion was the geodesicity,  in the sense of Weyl's invariant geodesics, of the 
 0-th order approximation of the current $j$ associated to the matter field. They proved that this criterion implies integrability of the length connection. Moreover the matter  $m$ associated to a  field propagates along the (invariant) geodesic by a factor $l^{-1}$ with 
\[ l:= e^{\int_{0}^{1} \varphi (\gamma ')ds  } \; , \]
the length transfer functionlike in our appendix I, equ. (\ref{length-transfer}).

In the light of  Dirac's version of the geodesic, the compatibility criterion implies only integrability of the length connection.  The subsequent argument of the  preferred choice of Riemann gauge given  in \cite{Audretsch_ea} is here no longer compelling.  Using the language of our appendix I, the authors showed that quantum mechanics requires calibration of mass transfer  with weight $-1$ along an  invariant geodesics (equ. 
(\ref{calibration-transfer})).  In other words,  the mass factor is   constant with respect to gauge covariant geodesics.  Audretsch/G\"ahler/Straumann showed, without realizing it,  that  {\em  Dirac's modification established just the right frame to make IWG  compatible with quantum mechanics.} In Dirac's approach the mass coefficients applied to  tangent vectors of timelike scale covariant geodesics are constant in any gauge.  
%%%%%%%%%%%
%%%%%%%%%%%%%%%%%%
 \\[0.5ex] \noindent
%\newpage \noindent
{\bf Gauge invariant   Einstein equation}\\ \noindent
In our approach  the   Weylian length connection   does not   introduce  a 
new field into the theory; it just adds another degree of freedom  to the 
metric,  the potential of the affine connection. Gravity continues to be uniquely described by the affine connection.  In such a weak  {\em extension of general relativity} (eGRT)  Weyl geometry  plays  a very modest 
role, in comparison to H. Weyl's original goal of formulating a unified field theory. But even  such a modest extension may be helpful for deepening  our understanding of the scaling invariance of gravity and electromagnetism and their interrelation. This  was already envisaged  in \cite{Canuto_ea}.  In most basic aspects our  approach  agrees with  this view, although we do {\em not use}  the ``large number principle'' (LNP) introduced by  Dirac in his studies of eGRT in the 1970s and taken over by Canuto and others.
Moreover, the innocent looking, but geometrically misleading  sign convention for gauge transformation has been corrected in the sequel.

The dynamics of eGRT remains governed by the Hilbert-Einstein action and the Einstein equation. 
Both have  to be reconsidered slightly in the framework of integrable Weyl geometry. As  scalar curvature $ \bar{R}$ is of  weight $-2$ and $\sqrt{|g|}$ of weight $4$, the Hilbert-Einstein action recquires an additional coefficient $\beta$, with a scalar scale covariant field $\beta$ of weight $-2$,  in order to form a gauge invariant Lagrange density:
\beq \label{Hilbert-Einstein-action} {\cal L}_{H} := \frac{ \beta}{16 \pi } \bar{R} \sqrt{|g|}   \; \eeq

We  assume a gauge invariant Lagrange density $ {\cal L}_M := L_M  \sqrt{|g|}$ of matter and 
non-gravitational fields, formed by a   scalar scale covariant field  $L_M$   of weight -4, complemented  by a vacuum Lagrange  density $ {\cal L}_V := C \beta ^2 \sqrt{|g|}$ with any free constant $C$. Variation of  
 the  first  and the last term of the combined action 
\[  {\cal S} := \int  (  {\cal L}_H  +  {\cal L}_M +    {\cal L}_V ) dx   \]
leads to 
\beqa
\frac{\partial {\cal L}_H}{\partial g^{\mu \nu }}\delta g^{\mu \nu }   &=&  \frac{\beta}{16 \pi } (Ric - \frac{1}{2} \bar{R} g _{\mu \upsilon } ) \sqrt{|g|} \delta g^{\mu \nu } \; \\
\frac{\partial {\cal L}_V}{\partial g^{\mu \nu }}\delta g^{\mu \nu } &=& C \beta ^2 g_{\mu \nu}  \sqrt{|g|}\delta g^{\mu \nu } 
 \eeqa

\noindent
If  we set
\beqa 
 T_{\mu \upsilon  }  &:=& - \frac{1}{2 \sqrt{|g|}} \frac{ \partial {\cal L}_M}{\partial g^{\mu \nu }} \;, \\
 \Lambda  &:=& 16 \pi  \beta C \, ,\eeqa
  the  Einstein equation   with  vacuum term
results:\footnote{Cf. \cite{Weinberg:Cosmology,Carroll:Spacetime} for the classical case and \cite{Canuto_ea} for the Weyl geometric variational characterization of the Einstein equation.}
 \beq \label{Einstein equation} Ric - \frac{1}{2} \bar{R} g = 8 \pi \beta^{-1} \, T  - \Lambda  g  \, 
 \eeq
The Weylian length connection is constrained by the condition
\beq \label{integrability}  d \varphi = 0 \; , \eeq 
as a necessary and sufficient for compatibility with quantum mechanics.  
Compatibility is here understood in the sense of  \citeasnoun{Audretsch_ea}. 

Obviously equations (\ref{Einstein equation}), (\ref{integrability}) are   scale gauge invariant. There is  no problem with the integrability condition (\ref{integrability}) or with the left hand side (l.h.s.) of the Einstein equation (\ref{Einstein equation}), which   contains  only scale invariant terms anyhow. 
 On the  right hand side (r.h.s.)  the gauge weights of the factors cancel:
\[ [[\beta^{-1}]] = 2 \, , \;\; [[T]] = -2 \, , \; \;  [[\Lambda ]] = -2 \, \;\; [[g ]] = 2 \, . \]
Because of dimensional considerations  Newton's gravitational constant  is of weight  $  [[N]]  = [[L^3]] [[T^{-2}]] [[M^{-1}]] = 3 -2 +1 = 2 $ and has to be characterized by a scale covariant field of weight $2$  in eGRT.  Thus the coefficient in equ. (\ref{Einstein equation}) can be identified  with it,
\[ \beta^{-1}   = N \,  .\]
 Then  the r.h.s. of (\ref{Einstein equation}) acquires the well  known form  $8  \pi N [\, c^{-4}]\, T - \Lambda  g$ 
 in physical units.\footnote{In general we use geometrical units with velocity of light  $c =1$. To facilitate the transition to empirical date we sometimes include  factors in powers of $c$ in square brackets. }

This small artifice allows to {\em transfer the classical Einstein equation into the context of integrable  Weyl 
geometry}. 

This procedure  makes  sense also from an empirical point of view. If we write $[L], [T], [E], [M]$ for the (physical) dimensions of length, time, energy, and mass respectively, and use  the relations $E = h \nu$ and $E = m c^2$ as  fundamental principles, we get for the gauge weights of the energy momentum tensor 
$ [[(T^i{}_j)]] = [[E]] + [[ L^{-3}]] = -4$; thus $[[(T_{ij})]] = [[g]] + [[(T^i_{}j)]] = 2 - 4 = -2 $.

From a field theoretical  point of view the extension by introducing  the connection $\varphi$ may look  trivial. In the next section it   will become clear  that $\varphi$ is nevertheless  able to  express   physical effects on long range electromagnetic radiation in the cosmos. It allows to treat space kinematical and other energy loss effects of photons on the cosmological level as mathematically equivalent. 
In addition we should not dismiss Riemann's  idea of fluctuations in the small which may cancel in medium range. It may lead to interesting results if applied to the scale
 connection.\footnote{Recently  Drechsler and Tann have started a research program which attempts to understand mass generation of quantum fields by oscillating integrable length connections \cite{Drechsler/Tann}.}
 %%%%%%%%%%
\\[0.5ex] \noindent
%\newpage \noindent
{\bf Atomic clocks and matter gauge}
\\ \noindent
 Recent high precision measurements give strong evidence that Newton's gravitational constant $N$ is  time independent  \cite{Will:LivingReviews}. A physically meaningful Weyl geometric extension of GRT has therefore to obey the following: 
\begin{principle} \label{clocks}
Measurements by atomic clocks   correspond  to a gauge in which the gravitational ``constant'', characterized by the scale covariant (scalar) field $N$,  is literally constant.  Such a gauge will be called  matter gauge. {\em (It is unique up to a global constant.)}
\end{principle}

In any full model of eGRT,  a scale covariant scalar field $\tilde{N}$ of weight 2 has to be specified, which defines the gravitational ``constant''. Starting from an arbitrary  representative of $\tilde{N}$ in a gauge $(\tilde{g},\tilde{\varphi})$ one easily   rescales   by $\Omega ^2 = C \tilde{N}^{-1}$
\[ g:= C \, \tilde{N}^{-1} \tilde{g}\, , \;\;\; \varphi := \tilde{\varphi} + \frac{1}{2}  d \log \tilde{N}\, ,\]
for any constant $C$. That leads  to {\em matter gauge}, because then $N = C\tilde{N}^{-1}\tilde{N}=C$. Predictions of the model have to be scaled in matter gauge, before they can be compared   meaningfully with empirical data taken by atomic clocks.

In this way, eGRT  expresses two well known insights of theoretical physics:
%%%%%%%%%%%%%%%%%
 \\[0.5ex] (1)  The field equations of gravity and electromagnetism are scale invariant.
\\[0.5ex] (2) The introduction of atomic clocks, or other metrically relevant matter structures, breaks the scale symmetry (``sponanteously'' as sometimes is added, although  with a slightly misleading field theoretic connotation).
%%%%%%%%%%%%%%%%%%
\\[0.5ex]
Classical relativity, in comparison, works with the {\em hidden postulate}
\[ \mbox{matter gauge} = \mbox{Riemann gauge} \, \]
inbuilt in its conceptual structure.
It breaks scale symmetry at an   unnecessarily early  stage of theory development.

From this perspective,  one may read Dirac's and P. Jordan's theories of a  ``time dependent gravitational constant''  as  first, unconclusive steps  towards a break with such an  identification (which they hesitated to do in the end). An  analysis of the Hubble effect (cosmological redshift) gives a stronger, even convincing reason to relax this traditional identification.
\\[0.5ex] \noindent
%\newpage  \noindent
{\bf Curvature quantities}
\\ \noindent
The  Riemann/Einstein gauge $(\tilde{g},0)$ of an integrable Weylian metric with gauge $(g,\varphi)$ may be considered as  a conformal  deformation of   the Riemannian component $g$ of  $(g,\varphi)$,  $ \tilde{g}  = \Omega ^2 g$. The Riemann and Ricci curvature tensors of the Weylian metric are gauge independent and  identical to those of $\tilde{g} $.   Ricci curvature $Ric = (R_{ij})$ of a Weylian metric $(g,\varphi)$  can therefore be calculated from the transformation formulae of  curvature quantities under conformal deformations as studied in 
\cite{Frauendiener:ConfInf} . In this way one finds:
\beq \label{Ricci IWG} R_{ij} = {}_gR_{ij} + 2 \varphi_i  \varphi_j + 2  {\nabla}_i \,  \varphi_j - g_{ij} \,(2  \varphi_l  \varphi^l - \nabla _l   \varphi^l ) \; 
\eeq
Here as elswhere in this article, front subscripts $_gX$ indicate the curvature quantity $X$ of the Riemannian component $g$ of the Weylian metric. $\nabla = (\nabla _i)$ denotes Weyl's {\em gauge invariant} derivative (the covariant derivative of the affine connection $\Gamma$ of the Weylian metric, cf. app. I, equ. (\ref{Christoffel})).

Taking the gauge weight -2 into account, the scalar curvature $\bar{R}$ becomes
\beq \label{scalar curvature IWG} \bar{R} =  {}_g\bar{R} - 6 \varphi_l  \varphi^l +  6 \nabla _l   \varphi^l  \; .
\eeq
The Einstein tensor $G$ in IWG  in the gauge $(g,\varphi)$   is
\beq \label{Einstein tensor IWG} G_{ij} =  {}_gG_{ij} + 2 \varphi_i  \varphi_j + 2  {\nabla}_i \,  \varphi_j + g_{ij} 
\,(  \varphi_l  \varphi^l - 2 \nabla _l   \varphi^l ) \; 
\eeq
\cite[equ. (13)]{Frauendiener:ConfInf}.\footnote{The   curvature formulae of conformal Lorentz geometry are helpful for correcting errors in the older literature on IWG. E.g., the expression for the Ricci tensor given in the otherweise very reliable contribution \cite{Canuto_ea} is wrong due to a sign error in one term, inherited from  \cite{Eisenhart:1926}. }

%\newpage
\subsection*{3.  Cosmology in  IWG}
 %\\[0.5ex] \noindent
{\bf Robertson Walker models}
\\ \noindent
The standard Friedmann-Lema\^{\i}tre approach explains  the Hubble effect, i.e. cosmological redshift,  by an  expansion of space sections.  Geometrically it  works with a  {\em 
Robertson-Walker manifold}  $M  = I \times_f S_{\kappa} $ over an open interval $I \subset \R$, with a Riemannian 3-manifold $ S_{\kappa}$ as standard fibre, which carries a metrical 2-form $d\sigma ^2$ of constant sectional curvature $\kappa$. $f$ indicates a   warp function (at least twice differentiable) $f: I \rightarrow \R^+$ for  the construction of the product metric
\beq  \label{R-W metric} \tilde g: \;\;    d\tilde{s}^2 = - [\, c^2]\, d \tau ^2 + ( f d \sigma )^2  \, . \eeq
A cosmological model is defined by   specifying  a timelike future oriented unit vector field
 $\tilde{X} $ as a {\em comoving observer field}. The natural choice
\[  \tilde{X} := \frac{\partial  }{\partial \tau }\]
leads to the well known {\em Friedmann-Lema\^{\i}tre models}  $(M, g, \tilde{X})$, if one imposes the Friedmann differential equation  as a constraint for $f$.

That translates easily to Weyl geometry. We only need to consider the observer field $X$   as a scale covariant field of weight $-1$, in order to make it unit in every gauge, $[[g(X,X)]]= 2-1-1 =0$. We shall then speak of a  Weyl-geometric {\em Robertson-Walker model}, generalizing the Friedmann-Lema\^{\i}tre ones. We denote it by $(M, [g, \varphi, X])$; the square brackets  indicate that the data of the Weylian metric and the observer field transform under scale gauges. Mathematically they are equivalence classes.\footnote{Compare, as a contrast, the   standpoint of treating classical  observer fields  in Weyl geometry (gauge weight 0) in \cite{Perlick:Observerfields}.}

We characterize the motion of photons semi-classically by a null-geodesic $\gamma $. Then  the photon energy $E(\tau )$ measured at a cosmic time $\tau $ in a comoving frame with time component $X(\gamma (\tau )))$ is according to the principles of GRT,
\beq E(\tau )  = g(\gamma '(\tau ), X(\gamma (\tau )) ) \; .\eeq
This expression  is obviously gauge invariant, $[[E]]=[[g]]+[[\gamma ']]+[[X]]=2-1-1=0$. Thus the {\em redshift} $z(\tau _0, \tau _1)$ during a transmission from a point $q_0$ to $q_1$, given by 
\beq  1 + z(\tau _0, \tau _1)  = \frac{E(\tau _0) }{E(\tau _1)} = 
\frac{g(\gamma '(\tau_0 ), X(\gamma (\tau_0 )) ) }{g(\gamma '(\tau_1 ), X(\gamma (\tau_1 )) )} \; ,  \eeq
is also gauge invariant. Our weak extension of GRT allows to modify the scale gauge without affecting the frequency shift of photons. That should be so in any reasonable physical usage of Weyl geometry, because frequency shift is measured by a dimensionless quantity (a ratio of quantities of the same type). 
%%%%%%%%%%%%%%
\\[0.5ex] \noindent
{\bf Einstein tensor for Robertson Walker models in IWG} 
\\ \noindent
%%%%%%%%%%%%%%
The Einstein tensor $G = (G_{ij})$ of a Robertson Walker Weyl model with gauge $(g, \varphi) $,
\[   g: \;\;    ds ^2 = - [\, c^2]\, d \tau ^2 + ( f d \sigma )^2 \; , \quad \varphi=(\varphi_0, 0,0,0) \; , \]
 is identical to the Einstein tensor of its Riemann gauged version, because of scale invariance of the Einstein tensor and is given by equation (\ref{Einstein tensor IWG}). It is obviously  diagonal. With $\varphi_0 =: H$, its  crucial components are:
\beqarr \label{R-W-W 00} G_{00} &=& \; _{g}G_{00} + 3 H^2 \; , \hspace{20mm} \quad  _{g}G_{00} = \frac{3}{f^2} (f'^2 + \kappa )\\
\label{R-W-W kk} G_{\alpha \alpha } &=& \; _gG_{\alpha \alpha } - g_{\alpha \alpha } (H^2 - 2 H') \; ,
\quad _gG_{\alpha \alpha }  = - \frac{1}{f^2} (2f''+f'^2 + \kappa ) g_{\alpha \alpha } \nonumber
\eeqarr
Like in the classical case,  the Einstein tensor $G$ has the form of the energy momentum tensor of an ideal fluid, 
\[   G_{00}  = 8 \pi N \rho  \; , \quad  \quad  G_{\alpha \alpha } = 8 \pi N p \; , \]
where the {\em energy density} $\rho $ and the {\em pressure} $p$ acquire additional terms from the Weylian length connection.

The {\em equation of motion} for $f$ is analogous to the classical case:
\beq  \label{equation of motion} 3(\frac{f''}{f}  -  H ')= - 4 \pi N (\rho +3 p)\eeq
The only difference lies in the additional term $-H'$ on the l.h.s..
%%%%%%%%%%%
%\newpage \noindent
\\[0.5ex] \noindent
{\bf Hubble gauge}
\\ \noindent
If in a gauge $(g, \varphi)$ of a model $(M, [g, \varphi, X])$ the redshift is given by the Weylian length transfer by the formula
\beq \label{Hubble gauge condition} 1 + z (\tau _0, \tau _1) = l (q_0, q_1) = e^{\int_{\tau _0}^{\tau _1}\varphi (\gamma ') d \tau }  \; , \eeq
we shall call $(g, \varphi)$ the {\em Hubble gauge } of the model. In such a gauge cosmological redshift is completely encoded by  the length connection $\varphi$. 

It is easy to verify that  in every Roberton-Walker model  a Hubble gauge exists. 
We only have to rescale the metric (\ref{R-W metric}) by $\frac{1}{f}$. That leads to the representation of the  Weylian metric by
\[g_1 = f^{-2} \tilde g:  \;\;    ds_1^2  = - \frac{d \tau ^2}{f^2} + d \sigma ^2  , \;\; \varphi_1 = - d \ln f^{-1} =   \frac{d f}{f} = \frac{f'}{f}d\tau. \]
Under such a rescaling, the natural (comoving) observer field $\tilde{X} := \partial_{\tau}$ of $\tilde{g}$ has to be rescaled   (with weight -1) to $X := f \tilde{X}$.  $X$ can then be expressed as a tangent field of  a coordinate time parameter $t$, if $f \partial_{\tau} = \partial_t \, \leftrightarrow f dt = d \tau $, or 
\[  \frac{dt}{d \tau}= \frac{1}{f} \, \; \leftrightarrow \; t = \int^{\tau}f^{-1} \, .\] 
This reparametrisation transforms the Riemannian component  of the Weylian metric  into a  static form, with the   length connection $\varphi =  \frac{f'}{f}d\tau = f' dt$:
\beq \label{Hubble gauge} g: \;\; ds^2  = - dt^2 + d\sigma^2 , \;\; \varphi = f'(t) dt \,  . \eeq
Here $\varphi = (\varphi^0, 0,0,0)$, $\varphi^0 = f'$,   contains all the information of the former warp function. 

The cosmological 
redshift of a photon is  known from standard cosmology
\[  1 + z(\tau _0, \tau _1)  = \frac{f(\tau _1) }{f(\tau _0)} \, .  \]
Because of \[  e^{\int_{\tau _0}^{\tau _1}\varphi_1 (\gamma ') d \tau} = e^{\ln f(\tau _1)- \ln f(\tau _0)}
=  \frac{f(\tau _1) }{f(\tau _0)} \; \]
condition (\ref{Hubble gauge condition}) is is satisfied. Thus  (\ref{Hubble gauge}) is the Hubble gauge of the Robertson Walker model. 

 The infinitesimal change of redshift at the cosmic time parameter $t $ is the Hubble ``constant'' (better Hubble function) $H(t)$ of the model in Hubble gauge
\[  H(t ) = \frac{df}{dt} = f'(t) \; . \]
It is different to the one seen in the standard approach,
 due to different scalings of the cosmological time parameter:
\[H_{stand} =  \frac{df}{d \tau} (\tau_0) = f'(t(\tau_0)) \, t'(\tau_0)  =   \frac{f'{(\tau_0)}}{f(\tau_0)} \; . \]

Photon energy $E$ with respect to an observer field is a scale covariant scalar field of gauge weight $[[E]] = -1$. Its motion according to Dirac's gauge covariant geodesics implies that  in Hubble gauge it propagates along invariant null geodesics  by calibration transfer of weight $-1$. 
The same holds for the wave vector $k$ of a photon along a null-geodesic and also for the mass of a freely falling particle.

Astronomical observations work  with related quantities of different gauge weight.  An important example ist the energy flux $F$ of the radiation  emitted by  astronomical sources. $F$ is observed as energy per time and per area. Its  gauge weight is $[[F]] = [[E]][[t]]^{-1}[[l]]^{-2}$ $ = - 1- 1-2  = - 4$. It seems natural to extend our first principle to such quantities by:
\begin{principle} \label{principle calibration transfer}
In Hubble gauge,  energy quantities of gauge weight $k$ transmitted by the compound system of the gravitational and electromagnetic fields propagate by  calibration transfer of weight $k$ in the sense of appendix I, equation (\ref{calibration-transfer}).
\end{principle}
  According to this principle, the  energy flux of astronomical sources decreases,  in Hubble gauge,  by a factor $l(\gamma (t))^{-4}$, where $l$ is  Weyl's length transfer function (\ref{length-transfer}). In the sequel  we shall use only this consequence. Readers who doubt the general principle  may put it back in second file.
%%%%%%%%%%%%%%
\\[0.5ex] \noindent
{\bf Extension of equivalence principle}
\\ \noindent
Let aus compare Riemann/Einstein gauge (\ref{R-W metric})  and Hubble  gauge  (\ref{Hubble gauge}) with repect to possible physical interpretations of our cosmological models. The first case suggests a physical {\em expansion of spatial sections} described by the warp function $f$. This interpretation is the default  choice of  present relativistic cosmology. It corresponds to the assumption that atomic clocks and energy tranfer of photons are naturally described by Riemann gauge (Riemann gauge is matter gauge, in our terminology). Equation (\ref{Hubble gauge}) shows that the same properties of cosmological redshift can be described equivalently  in a geometry respecting the principles of (weakly extended) GRT, but  without any contribution from space expansion or  diverging  flow lines of the observer field.   Hubble gauge (\ref{Hubble gauge}) may even be physically `true', i.e. adequate, if cosmological redshift is due to an {\em energy loss of photons} during the passage of long distances in the compound system of the cosmic vacuum and the background electromagnetic and gravitational  fields. 

F. Zwicky and other representatives of the first generation of relativistic cosmologists, among them E. Hubble, R. Tolman and H. Weyl, assumed such a field theoretic, ``more physical'' cause \cite[300]{Weyl:Redshift} for cosmological redshift. In the second half of the 20th century this assumption was revived  under the name of  ``tired light'' hypothesis (J.P. Vigier e.a.). The  specific version of Vigier's tired light hypothesis was refuted by astronomical observations in the 1970s, not the general assumption behind it.

 In the Weyl geometric framework space expansion and energy loss in the ether become mathematically equivalent expressions for the same cosmological redshift,  if we  use,   like Einstein, de Sitter, Weyl and others, the abbreviated term {\em ether}  for the compound system of background fields (and the quantum vacuum, we have to add now). In this sense, eGRT allows to {\em   extend Einstein's equivalence principle in a mathematically  precise form to space kinematical effects and energy loss of photons} in the cosmic ether.  
%%%%%%%%%%%%
%\newpage  \noindent
\\[0.5ex] \noindent
{\bf Affine connection}
\\ \noindent
Using Riemannian coordinates $x_{\alpha }$  in the spatial fibre $S_{\kappa}$ the spatial part of the metric is
\beq  \label{Riem coordinates} d \sigma ^2  = \frac{\sum_{\alpha =1}^3  dx_{\alpha }^2 }{ (1+ \frac{\kappa }{4}  \sum_{\alpha =1}^3 x_{\alpha }^2 )^2 } \,  .\eeq  
With the abbreviations
\beq \label{B} B :=    (1+ \frac{\kappa }{4}  \sum_{\alpha =1}^3  x_{\alpha }^2 )^{-1} \; , \quad \;\; H:= f' \; \eeq
the non-vanishing coefficients of the affine connection of a Hubble gauged Robertson-Walker model with coordinates as in (\ref{Hubble gauge}) become (Greek indices $\alpha ,\beta, \gamma = 1,2,3$, Latin indexes $i, j, k = 0, \ldots, 3$):
\beqarr \label{affine connection} \Gamma ^0_{00} &=& [c]\, H \, , \quad \Gamma ^{\alpha }_{0\alpha } = [c]\, H \, ,
  \; \quad \Gamma ^{0 }_{\alpha \alpha } = B^2 \frac{H^2 }{[c]}  , \\ \nonumber
 \Gamma ^{\alpha }_{\alpha \alpha } &=&   \Gamma ^{\beta  }_{\alpha  \beta  } = - B \frac{\kappa  }{2} x_{\alpha } \, , \quad
 \Gamma ^{\beta  }_{\alpha \alpha   } =  B \frac{\kappa  }{2} x_{\beta  } \, , \quad \mbox{for  } \;\; \beta  \not= \alpha .
\eeqarr
Although the affine connection is gauge invariant the expression of $\Gamma$ differs from the ordinary  Christoffel symbols $\tilde{\Gamma } $ of Robertson-Walker manifolds (Riemann gauge), because of the different parametrizations of  cosmological time. A comparison shows that  $\tilde{\Gamma}^0_{00} = 0$, while all the other Christoffel symbols look formally the same as in equ. (\ref{affine connection}). This is also the case for the two other components of the affine connection, containing cosmological  terms, $\tilde{\Gamma} ^{\alpha }_{0\alpha }  = [c]\, H$ and $ \tilde{\Gamma} ^{0 }_{\alpha \alpha } = B^2 \frac{H^2 }{[c]}$.    But in this context the Hubble function is different, $H = \frac{f' }{f}$.  For possible dynamical consequences for low velocity orbits only the difference in  $\Gamma ^0_{00} $ matters.  $H$ is equal to the observational value of the Hubble constant $H_0$ anyhow, and in this sense independent of its representation in the model.
%%%%%%%%%%%%%%
\\[0.5ex] \noindent
{\bf Consequences for low velocity orbits}
\\ \noindent
Dynamical consequences of cosmological contributions to the affine connection can be calculated by   {\em post-Newtonian approximation} to geodesics (scale covariant or invariant) \cite[213ff.]{Weinberg:Cosmology}. We only have to apply the method  to integrable Weyl geometry and eGRT.\footnote{\label{fn Maeder}Nearly 30 years ago A. Maeder   calculated low velocity  approximations  of Weyl geometric geodesics    in the Dirac-Canuto  research program  by essentially the same method 
\cite{Maeder:LowVelocity}. Our understanding of  eGRT  demands some seemingly minor  modifications. Interestingly enough, they have   decisive  consequences.}

The equation of motion for mass points in eGRT, parametrized  in coordinate time $t$ is (see appendix II,  equ. (\ref{equ of motion})):
\beqa 
\frac{d^2 x^{\alpha }}{dt^2} &= & 
- \Gamma^{\alpha}_{00} +\Gamma^{0}_{00} \frac{dx^{\alpha}}{dt}
 - 2 \Gamma^{\alpha}_{0 \beta} \frac{dx^{\beta}}{dt}
-  \Gamma^{\alpha}_{ \beta \gamma} \frac{dx^{\beta}}{dt}\frac{dx^{\gamma}}{dt}  \\
& & + 2  \Gamma^{0}_{0 \beta } \frac{dx^{\alpha}}{dt}\frac{dx^{\beta}}{dt}
+ \Gamma^{0}_{ \beta \gamma } \frac{dx^{\alpha}}{dt}\frac{dx^{\beta}}{dt}\frac{dx^{\gamma}}{dt}
\eeqa
 This is formally identical to the result in Riemannian geometry \cite[equ. (9.1.2)]{Weinberg:Cosmology}, but here the affine connection is the one of Weyl geometry.

We consider a gauge $(g, \varphi )$  of a Weylian metric with  orthogonal coordinates, $g = diag(g_{00}, g_{11},g_{22}, g_{33})$ and  $\varphi$ of  the simple form $(\varphi_0, 0,0,0)$ as in all our cosmological applications (we shall use the terminology $\varphi$ is {\em  cosmological}). 
The Christoffel symbols which are  crucial for low velocity approximations up to first order velocity terms can be read off from equ. (\ref{Christoffel}):
\beqa  \Gamma^{0}_{00} &=& {}_g\Gamma^{0}_{00} + \varphi_0\, , \; \; \;  
\Gamma^{\alpha}_{0 \alpha} = {}_g\Gamma^{\alpha}_{0 \alpha}   + \varphi_0 ,  \; \; \;   \\ \Gamma^{\alpha}_{00} &=& {}_g\Gamma^{\alpha}_{00}  \, ,  \; \; \;   
\Gamma^{\alpha}_{0 \beta} = {}_g\Gamma^{\alpha}_{0 \beta} \;\;  \mbox{for } \alpha \not= \beta \, ,  \; \; \;    \Gamma^{\alpha}_{ \beta \gamma} = {}_g\Gamma^{\alpha}_{ \beta \gamma}  
\eeqa

That leads to the  equation of motion 
for low velocities: 
\[   \frac{d^2 x^{\alpha }}{dt^2}  \approx  
- \Gamma^{\alpha}_{00} +\Gamma^{0}_{00} \frac{dx^{\alpha}}{dt}
 - 2 \Gamma^{\alpha}_{0 \beta} \frac{dx^{\beta}}{dt}  \]

\beq \label{lv equation} \frac{d^2 x^{\alpha }}{dt^2}  \approx  
- {} _g\Gamma^{\alpha}_{00} + {}_g\Gamma^{0}_{00} \frac{dx^{\alpha}}{dt}
 - 2 \, {}_g\Gamma^{\alpha}_{0 \beta} \frac{dx^{\beta}}{dt} - \varphi_0  \frac{dx^{\alpha}}{dt}
\eeq

The last term of the equation contains the difference between the coordinate acceleration of dynamics defined by the semi-Riemannian component alone  and the Weyl geometric modification. It shows that  {\em  a non vanishing cosmological Weylian  length connection $\varphi$ leads to an additional coordinate acceleration $a_H$ proportional to the velocity, but of inverse sign,  with $\varphi_0$  as  coefficient}:\footnote{\label{fn Maeder 2}A similar result  was derived by A. Maeder, but with a different  (wrong) sign. Due to a sign inconsistency between scaling function and length connection, inherited from 
the reading of Weyl's gauge transformation common in the physical literature, Maeder was misled to believe that the Hubble effect of Weyl geometry leads to  a (positive)   acceleration and may explain the cosmic ``flight'' tendency leading to the Hubble redshift \cite{Maeder:LowVelocity}. It is the other way round; the cosmological terms of the guiding field induce a  {\em  deceleration} for low velocity orbits and is  a {\em consequence} of the Hubble effect.  Compare appendix III. } 
\beq  \label{cosmic deceleration IWG} a_H = - \varphi_0   \frac{dx^{\alpha}}{dt} = - H \frac{dx^{\alpha}}{dt} 
 \eeq
In the Hubble gauge of Robertson-Walker manifolds $\varphi_0 = H$ is the Hubble  function (in the case of  Weyl universes below even constant, $H = H_0$).

This result has to be compared with the one derived in  the corresponding Friedmann-Lema\^{\i}tre model  (Hubble gauge versus Riemann 
gauge). 
For the latter the cosmological correction term of coordinate accelerations is 
\beq \label{cosmic deceleration F-L} a_H = - 2 H_0 \frac{dx^{\alpha }}{dt} \; ,  \quad \mbox{with }  \quad  H_0 := \frac{f'}{f}(t_0)\; \;\eeq 
(equ. (\ref{lv Friedmann}), app. III).
 The cosmological acceleration terms  of the standard Friedmann-Lema\^{\i}tre approch and the Weyl geometric models are formally analogous, but differ by  a factor 2. This  difference has  observable consequences.
%%%%%%%%%%%%%%%%
\\[0.5ex] \noindent
{\bf Solar system observations}
\\ \noindent
The observational value for the  the Hubble constant is
\[ H_0 \approx 2.27 \, (\pm 0.2) \, 10^{-18} \; s^{-1} \]
corresponding to $ H_0 \approx  70\,\; km s^{-1} Mpc^{-1}\;  (\pm 8 \%)$.  With typical velocities of spacecrafts   like Pioneer 10 or 11, $v \approx  3 \cdot 10^6 \; cm \, s^{-1}$,  we have to expect additional decelerations from the cosmological correction term for such velocities at the order of magnitude
\[  a_H = - H_0 v \sim  10^{-12} \; cm \, s^{-2} .\]
This is 9 orders of  magnitude below a typical value of gravitational acceleration $ a_{10AU} $ in the solar system at the distance of $10$ astronomical units $AU$, and  4 orders of magnitude below the anomalous acceleration  $a_P$ of the Pioneer spacecrafts determined in the late 1990s \cite{Anderson_ea:Pioneer_I},
\[ a_{10AU} = \frac{N M }{(10 AU)^2}  \approx 5.9 \;10^{-3}\; cm \,s^{-2} \; ,  \quad  a_p \approx 8.74\, (\pm 1.33)\; 10^{-8} cm \, s^{-2} \; \]
($N$ Newton constant, $M$ solar mass).

 Present solar system tests of GRT work at an error marge corresponding to acceleration sensitivity several orders of magnitude larger than $a_H$  \cite{Will:Book,Will:LivingReviews}.
With respect to this evidence eGRT in our sense is effectively an $\alpha_1 =\ldots = \alpha _4 = \zeta _1=\zeta _2=\zeta _3 = 0,  \beta = \gamma  = 1$ theory  in terms of the parameters $ \alpha _1, \ldots, \zeta _1, \ldots , \beta, \gamma  $ of  parametrized postnewtonian gravity (PPN). By present standards the cosmological correction of eGRT cannot be distinguished from classical (semi-Riemannian) relativity on the level of solar system observations. If it could, also the  cosmological terms of the affine connection in the standard approach would lead to  observable consequences. 

This observation does not  exclude the possibility that other modifications of the Weylian scale connection, arising locally and differing from the large cosmological mean, may have observable consequences inside  the solar system.
%%%%%%%%%%%%%%%%%%
%\newpage
 \\[0.5ex] \noindent
{\bf Determination of dynamical mass}
\\ \noindent
For the determination of dark matter by dynamical observations  of large mechanical systems on the cluster or super cluster level (virial theorem or generalizations) cosmological correction terms for the dynamical equations of point masses have to be taken into account.\footnote{Cf. \cite[581f.]{Peebles:Constant}, more in detail \cite[63ff.]{Peebles:LargeScale}.} J. Peebles looks at   velocities $\vec{v}$ and accelerations $\vec{g}$ of galaxies ``relative to the homogeneous background model'', i.e., relative to a kind of  `comoving' observer system {\em without} expansion (he speaks of  ``peculiar'' velocities). They are different  from the observed velocities and accelerations, which   include cosmological effects. By a striking heuristic argumentation Peebles introduces a cosmological correction term in an expanding universe model with expansion function $a(t)$ and arrives at at total acceleration of the form:
\beq \label{Peebles corr} \frac{\partial \vec{v}}{\partial t} = \vec{g} - \frac{\dot{a}}{a} \vec{v} \quad \quad \mbox{\cite[equ. (51)]{Peebles:Constant}
 } \eeq
As $\frac{\dot{a}}{a} (t_0)=H_0$,  this comes down to using a {\em cosmological correction term which is consistent with the  low velocity dynamics in IWG (equ. (\ref{cosmic deceleration IWG})) and inconsistent with 
Friedmann-Lema\^{\i}tre models (equ. (\ref{cosmic deceleration F-L}))} or Riemann/Einstein gauge of Robertson Walker manifolds in general (compare app. III). 

The recent investigations of mass densities by different methods have led to comparably sharp constraints for the mass density in the universe.  Peebles gives intervals with factor $\leq 3 $ between upper and lower bound, $0.15 \leq \Omega _{m} \leq 0.4$ or even  $0.15 \leq \Omega _{m} \leq 0.3 $ \cite{Peebles:Constant,Peebles:Probing}. Already on the cluster level the cosmological correction term is at the order of magnitude of the ``peculiar'' acceleration (in Peeble's terminology).\footnote{\cite[84]{Maeder:DynamicalMass}} Its substitution by the cosmological correction term $-2H_0 \dot{v}$ of the Friedmann-Lema\^{\i}tre approach   would  lead to considerable deterioration of data coherence. For super cluster data we have to expect a complete breakaway from the factor smaller 3 achieved during the last years.

By obvious reasons we  demand mathematical 
consistency between evaluation procedures of empirical data and the cosmological model accepted as theoretical background frame. 
The evaluation of observational data of dynamical mass according to J. Peebles' cosmological correction term is consistent with IWG, rather than with the Riemann/Einstein gauge presupposed in F-L models.  
   {\em Thus the present dynamical data on mass densities and their comparison with other methods speak strongly in favour of   the Weyl geometric approach to Robertson-Walker manifolds.} We have to consider  Hubble gauge as a serious candidate for defining the physical metric (matter gauge in our terminology). 
We therefore turn towards investigating the simplest cases of Weyl geometrical models  under the premise of Hubble gauge as matter gauge.

\subsection*{4. Weyl universes}
 
We start by studying  
the simplest models of Weyl geometrical Robertson Walker cosmologies with constant Hubble connection,  $f'  = const =: H$,   $\varphi = H dt$. They are time homogeneous,  behave nicely  from the physical point of view  ( in Hubble gauge) and have good properties as cosmological  models. 
\\[0.5ex] \noindent
{\bf Definition}
\\ \noindent
More precisely, we  work in a manifold  $M  = I \times S_{\kappa} $ with Weylian metric
\beq \label{Weyl universe} g: \;\; ds^2 = - [\,c^2]\, dt^2 + d\sigma^2 \; ,\; \; \varphi = H dt  \, ,\eeq
where $H$ is a  constant in the literal sense (usually denoted by $H_0 :=H$). $ d\sigma^2$ is  the metric on a space of constant sectional curvature $S_{\kappa}$,  given, for example, in Riemann coordinates    (equ. \ref{Riem coordinates}).
Such a model  will be  called a {\em Weyl  universe}. (\ref{Weyl universe}) is its Hubble gauge. 
We shall speak of an {\em Einstein-Weyl} universe in the case $\kappa >0$, of {\em Minkowski-Weyl} or {\em Lobachevsky-Weyl} universes for $\kappa =0$ or $\kappa <0$, respectively. A Weyl universe will be called 
 {\em non-degenerate}, if $H > 0$.
For the sake of simplicity,  we  assume simply connected spatial fibres and the validity of the Poincar\'e conjecture in dimension 3  (which seems  to be proven).

 In addition to the empirical argument given above, the properties of the r.h.s. of the Einstein equation give strong theoretical reasons  to demand the scale covariant field   $N$ to be constant in this gauge (principle \ref{clocks}).   Hubble gauge  is  the most natural candidate for matter gauge in these models (see below, energy momentum tensor).
%%%%%%%%%%%%%%
\\[0.5ex] \noindent
%\newpage \noindent
{\bf Redshift}
\\ \noindent
In a Weyl universe the length transfer function of equ. (\ref{length-transfer})  along any path $\gamma (s) = (s, c(s))$ with $c(s) \in S_{\kappa }$  and $\gamma ' = u$ from an event $q_0$ lying above $t_0$ to an event $q_1$ above $t_1$ is
\beq \label{length transfer Weyl universe}  l(q_0,q_1):= e^{\int_{t_0}^{t_1} \varphi (u)} =  e^{\int_{t_0}^{t_1} H ds}= e^{H (t_1 - t_0)} \, .  \eeq  
 Because of path independence it is identical   to the integral  along a lightlike curve,  thus $1+z= e^{H (t_1 - t_0)} $. The  distance in gauge (\ref{Hubble gauge}) between two events $q_0 = (t_0,p), q_1=(t_1, p)$  with the same projections $p$ in the spatial fibre $S_{\kappa}$,  is 
$ d(q_0,q_1) = [c] ( t_1 - t_0) .$ For events $q_0 = (t_0, p_0), q_1=(t_1,p_1), p_0, p_1 \in S_{\kappa }$, with lightlike connecting path and of ``spatial distance'' $d(p_0,p_1)$ in $S_{\kappa }$, the cosmological redshift from $q_0$ to $q_1$ is 
\beq \label{redshift Weyl universe} 1+z =e^{H (t_1 - t_0)}  = e^{H_1 d(q_0,q_1)}\, , \; \;  \; H_1 := c^{-1} H= c^{-1} H_0 \, .     \eeq
%%%%%%%%%%%%%%
%\\[0.5ex] \noindent
{\bf Semi-Riemannian pictures}
\\ \noindent
If one wants to  consider a Weyl universe from the semi-Riemannian point of view, one has to use the length transfer function of equ. (\ref{length transfer Weyl universe}) as  scaling factor,
\beq  \label{exponentially expanding metric} \bar{g}:   d\bar{s}^2 = e^{2 H (t_1 - t_0)} (- dt^2 + d\sigma) \, .   \eeq
This metric is being discussed in the literature under the name {\em scale expanding cosmos} for the special case $\kappa = 0$ (more precisely, scale expanding Minkowski space)  \cite{Masreliez:Scale_Expand}. 
Taken at face value, $\bar{g}$ does {\em not} define a proper cosmological model, as the observer field associated to the coordinate time $X = \partial t$ is not unit. It has the norm $|X|^2 = - e^{2Ht}$. One may  choose between two possibilities  of ``rescaling''.

 Rescaling the  coordinate time only by $\tau := H^{-1} e^{Ht} \leftrightarrow t = H^{-1} \log H \tau$ leads to a {\em linearly expanding  Robertson-Walker manifold}  $M = \R \times_{f} S_{\kappa }$. It is just another description of the metric (\ref{exponentially expanding metric}) in Riemann gauge:
\[g: \;\; ds^2 = - dt^2 + (H t)^2 d\sigma^2  \, \]
Rescaling  the  whole metric  by the factor $\Omega (t)^2 = e^{-2Ht}$ leads back to Hubble gauge,  equ. (\ref{Weyl universe}), of   the {\em Weyl universe}, here with flat ex-ante space sections 
(i.e. the  Minkowski-Weyl universe).\footnote{J. Masreliez argues with the unobservability of scale expansion for material measurements. In this way he implicitly  uses Hubble  gauge for the comparison of theoretically derived quantities with observational values. He does not  draw, however,  full advantage of the underlying Weyl geometric structure.}
%%%%%%%%%%%%%%
\\[0.5ex] \noindent
{\bf Isometries}
\\ \noindent
Two Weyl universes are isometric if their ratios $\frac{\kappa  }{H^2}$ coincide \cite[Prop. 3]{Scholz:Extended_Frame}. We therefore introduce the {\em module} of a Weyl universe
\beq \label{module} \zeta := \frac{\kappa  }{H^2} \; .\eeq 
The non-degenerate Weyl universes ($H > 0$) form a continuous 1-para\-me\-ter space of isomorphy classes, characterized by the module $\zeta $.
The Hubble constant $H$ is a (global) scaling 
quantity. Empirically it has  been measured with great accuracy, in comparison with other cosmological 
data. Thus the class of Weyl universes has  only one essential metrical parameter which has to be fitted and compared with observational values. The supernovae data indicate clearly that  in our model class $\zeta > 0$, in fact it is at the order of magnitude $ \zeta \sim  1$   (see end of section 5).   In the sequel we therefore concentrate on 
{\em Einstein-Weyl} models.
 %%%%%%%%%%%%%%
\\[0.5ex] \noindent
{\bf Dynamical deformations} 
\\ \noindent
For the investigation of stability questions we have to  leave the restricted class of Weyl universes and  consider metrical deformations which keep maximal symmetry of the spatial fibres and the Hubble connection. A generic exemplar of a dynamically deformed (gauged) Weyl universe (\ref{Weyl universe}) is given by $M = I \times _f S_{\kappa }$ and a Weylian metric 
\beq  \label{deformed Weyl universe}  g: \;\; ds^2 = - [\,c^2]\, dt^2 + f^2 d\sigma^2 \; ,\; \; \varphi = H dt  \, ,  \eeq
with twice differentiable function $f: I \longrightarrow \R^+$, close to the constant function $f_1 \equiv 1$ and constant $H$. Equation (\ref{equation of motion}) then reduces to the classical equation of motion for the warp function of  Robertson-Walker universes:
\beq  \label{equation of motion Weyl universes} 3\frac{f''}{f}  = - 4 \pi N (\rho +3 p)\eeq
We come back to this equation  in section 5 when we discuss  stability questions.
%%%%%%%%%%%%%%
\\[0.5ex] \noindent
{\bf Curvature}
\\ \noindent
The affine connection of the Hubble gauged Robertson Walker models (equ. (\ref{affine connection})) leads to the following curvature data of Weyl universes (compare also equs. (\ref{Ricci IWG}), (\ref{scalar curvature IWG}), (\ref{Einstein tensor IWG})): 
\\[0.5ex] \noindent
 {\em Riemann tensor}
\beqa  R^{\alpha} _{\beta \alpha \beta } &=& (\kappa + H^2) B^2 \quad \mbox{for } \; \alpha \not= \beta \, , \quad\quad \mbox{$B$ as in equ. (\ref{B}), }   \\
  R^i_{jkl} &=& 0 \quad \quad \quad \quad  \quad \quad \mbox{in all other cases, } \;\;
 \eeqa
{\em Sectional curvatures }
\beqa  \kappa _S &=& \kappa + H^2   \quad \mbox{ in 2-directions tangential to  the spatial fibres } \;\; \\
 \kappa _T &=& 0 \quad  \quad \quad \quad \mbox{in surface directions containing a timelike vector, } \;\;
\eeqa
{\em  Ricci tensor }
\beqa Ric = 2 (\kappa +H^2) d\sigma ^2 \; , \;  \quad R^i_j = 2(\kappa +H^2)\; diag (0, 1,1,1) \; ,
\eeqa
 {\em Scalar curvature }
\[  \bar{R} = 6(\kappa +H^2) \; .\]
The curvature quantities of Weyl universes look  exactly like those of a static model  with spatial sectional curvature $\kappa ' = \kappa + H^2$. The Hubble connection seems to supply the spatial fibres with an additional curvature term $H^2$. That explains the terminology {\em ex-ante} sectional curvature for $\kappa $ and  of {\em total}, or {\em effective} sectional curvature for $\kappa + H^2$ (of the space fibres). Note that the affine connection behaves differently (equ. (\ref{affine connection})). In spatial directions it is equal to  the one of static cosmological models, but it has additional terms in timelike directions.

Historically minded readers may like to re-read Milne's ``kinematical cosmology'' with  linearly expanding spatial fibres in Minkowski space as the Robertson-Walker picture  of a Lobachevsky-Weyl universe with $\zeta =-1$  \cite[60ff.]{Kragh:Cosmos}. The more recent ``scale expanding cosmos'' is an example of a Minkowski-Weyl space ($\zeta = \kappa =0$). We have reasons to prefer the Einstein-Weyl case ($\kappa > 0$).
%%%%%%%%%%%%%%
\\[0.5ex] \noindent
%\newpage \noindent
{\bf Einstein tensor}
\\ \noindent
The l.h.s  of the Einstein equation of a Weyl universe can now directly  be read off. In Hubble gauge it is
\beq  \label{Einstein-tensor}  Ric - \frac{1}{2}\bar{R}g = 3 ( \kappa + H^2) dt^2   -   ( \kappa + H^2) d\sigma^2 \; . \eeq
It presupposes a r.h.s. $ 8 \pi N \, T $ with an energy stress tensor $T$ of an ideal fluid with total energy density $\rho $ and negative pressure $p < 0$, satisfying the ``strange'' equation of state,
\beq \label{hyle} p = - \frac{1 }{3} \rho  \; ,  \eeq
like in the classical static models.
We refer to it as the {\em hyle equation}. Here the word {\em hyle} is used as a metaphorical expression for the compound materio-energetic system which lies at the base of the r.h.s. of the cosmic Einstein equation. 
Its pressure is related to the geometrical parameters by 
\beq \label{pressure rhs} p =  - \frac{ \kappa + H^2 }{8 \pi \, N} [\, c^4] = -\frac{H^2}{8 \pi N}(\zeta +1) [\, c^4]  \; .\eeq

 Clearly energy density and pressure are   {\em constant in Hubble gauge}. Although  constant solutions have been considered as unrealistic since the detection of cosmological redshift in the 1920s,  we are no longer forced to so in the light of the Weyl geometric approach. 

 In Riemann gauge energy density depends on the cosmic time parameter  $\tau $ in such a manner ($\sim \tau ^{-4}$) that a reasonable physical interpretation seems to be excluded. The total energy content of space sections is not constant in this gauge and  the Friedmann equation does not hold \cite[351]{ONeill}.
 If we want to investigate whether Weyl universes may be useful for modelling cosmic geometry, {\em we have to accept Hubble gauge} as a reasonable candidate for characterizing the behaviour of atomic clocks (principle \ref{clocks}). 
%%%%%%%%%%%%%%
%\\[0.5ex] \noindent
\newpage \noindent
{\bf $\Lambda  $ term}
\\ \noindent
For $\Lambda  \not=  0$,  $T$ contains a contribution  proportional to  the term $-\Lambda  g$ in equation (\ref{Einstein equation}), in addition to better known forms of  radiation energy and matter content of the universe, which we abbreviate by $T_{rad}$ and $T_m$:
\beq \label{T tilde}  T  = T_m +  T_{rad}  - \frac{1 }{8 \pi \, N}\Lambda  g = T_m +  T_{rad} + T_{\Lambda  }  \; , \eeq
where $T_{\Lambda  } =  - \frac{1 }{8 \pi \, N}\Lambda  g $ is ascribed to the {\em cosmic   vacuum}. 

 If  $T$ is  solely due to dilute, cold (pressure free) matter, $T_{rad}= 0$, we get $  T_{\Lambda  } =   \frac{ \kappa + H^2 }{8 \pi \, N}\, (dt^2 - d\sigma ^2)) $. The  energy densities $\rho _m$ and $\rho _{\Lambda  }$ of $T_m$ and $T_{\Lambda  }$ become
\[ \rho _m = \frac{ 2 ( \kappa + H^2) }{8 \pi \, N}[\, c^4] \quad \mbox{and } \;\;\rho_{\Lambda  }  =  
 \frac{ \kappa + H^2 }{8 \pi \, N}[\, c^4] \; . \] 
Under this  special assumption,  much too restricted as we shall see in section 5, the energy densities  have  relative values similar to those of the classical Einstein universe,
\[  \Omega   _m := \frac{\rho _m }{\rho _{crit}}  = \frac{2 }{3} (\zeta +1) \; , \quad \quad 
\Omega _{\Lambda  }:=  \frac{\rho _{\Lambda  } }{\rho _{crit}} =  \frac{\Omega _{m}}{2}\; ,\]
where 
 \beq \label{critical energy density} \rho _{crit} = \frac{3}{8 \pi \, N}   H^2  [\, c^4] \; \eeq  
 is the  the critical  energy density of the standard approach.

For $T_{rad} \not= 0$, the state equations of the Maxwell field
$ p_{rad} = \frac{1}{3} \rho_{rad} \;   $% \label{state Segal} 
and of the vacuum
$ p_{}\Lambda = - \rho _{\Lambda } $
lead to the pressure condition
\[  \rho _{\Lambda } - \frac{1}{3} \rho _{rad}  = \frac{\kappa + H^2}{8 \pi N} \; .\]
 Therefore the equations
 (\ref{T tilde}), (\ref{Einstein-tensor}) teach us that:
\beqarr \Omega _m + \Omega _{Seg} + \Omega _{\Lambda  } & =& \zeta  + 1 \quad \quad 
\mbox{(energy density condition)}  \label{energy density condition}  \\ % 
\Omega _{\Lambda  }  - \frac{1}{3} \Omega _{Seg} &=&  \frac{\zeta +1}{3} \quad  \quad \mbox{(pressure condition) } \;\; \label{pressure condition}%
\eeqarr
%%%%%%%%%%%%%%
\\[0.5ex] \noindent
{\bf Light cone and light spheres}
\\ \noindent
Because of conformal invariance of the light cone, the  traces of null geodesics  in Weyl universes with Weyl gauge $(g, \varphi)$ are identical to traces of null geodesics of the Riemannian component $g$ only. They can easily be described. Let us denote the length of the projection of the segment $t \leq s \leq t_0$ of a  null geodesic $\gamma (s)$ into the spatial standard fibre $S_{\kappa }$  by $r$. Abbreviating the redshift between $t$ and $t_0$  by $z$, we find
\[  r= c |t - t_0| = H^{-1} \ln (z+1) . \]
We  call    the  intersection of the light cone with spatial fibres  a {\em light sphere}  and $r$ its  {\em radius}. 

%\begin{lemma}\label{lemma light-spheres}
 In a Weyl universe of ex-ante sectional kurvature $\kappa $, Hubble constant $H$, and module $\zeta = \kappa H^{-2}$,   the area 
 $O = O_\kappa $ of   light spheres in dependence of redshift  can easily be calculated. It is 
\[  O_\kappa = \frac{4 \pi}{\kappa } SIN_k^2 (\sqrt{  \zeta } \ln (z+1)  \;  , \]
 where $k = \pm 1$ or $0$ and $SIN_1 := \sin$, $Sin_0 := id$, $SIN_{-1} := \sinh $.% \\[0.8pt] \noindent

The volume $V(z_1, z_2)$ scanned by the light cone between redshift values $z_1 \leq z_2$ is accordingly\footnote{The  proof of the first statement is obvious, as we calculate  areas of  spheres with  radius $r = H^{-1} \ln (z+1) $ in   the  3-geometries of constant curvature radius $a = \kappa ^{-\frac{1}{2}}$   and use (\ref{module}). For the volumes   we only need to integrate over  infinitesimal volume layers $O d x^0$ of the light cone, substitute  $d x^0 = c d |t|$, and use equation (\ref{redshift Weyl universe}). }  
\beq \label{volume}   V(z_1, z_2) = \int \limits_{z_1} ^{z_2} (z+1)^{-1} O(z) dz  .\eeq
%%%%%%%%%%%%%%
\\[0.5ex] \noindent
{\bf Apparent luminosities of astronomical sources}
\\ \noindent
Because the energy flux   $F(z)$  of a radiating source propagates by calibration transfer of weight $-4$ (principle
 \ref{principle calibration transfer}) and distributes over light spheres with area $O(z)$, it depends on redshift like\footnote{Readers who are doubtful of principle \ref{principle calibration transfer},  may want  to avoid  it and prefer to  calculate  in the Robertson Walker picture 
 with traditional methods of scaling down the flux in  ``expanding universes'' (not to forget the gauge factors between Weyl and Riemann gauge).  The result is the same.   }
\[  F(z)  \sim   (z+1)^{-4}   O(z) ^{-1}  \; .\]
After redshift  correction of the measured energy by the factor $(z+1)$, the energy flux $F_{corr}$  is proportional to
\[  F_{corr}  \sim (z+1)^{-3}   O(z) ^{-1}   .\]
Then the luminosity distance $d_L^2 := \frac{L}{4 \pi F}$, with $L$ absolute luminosity and $F$ the measured flux becomes:
\beq \label{luminosity distance} d_L \sim  \frac{(1+z)^2}{\sqrt{\kappa }} SIN_{k } (\sqrt{\zeta } \ln (1+z))   \eeq
%%%%%%%%
For the apparent magnitude of a cosmic source, 
\[  m := - 2.5 \log F  + C , \]
where the constant $C$ contains the dependence on the  absolute magnitude $M$,
  we get
\[   m = - 2.5 \log (z+1)^{-3} O(z)^ {-1}  + C = 5 \log (z+1)^{\frac{3}{2}} \sqrt{O(z)} + C \; ,  \]
or
\beq \label{luminosity}
m_k(z)  = 5 \log \left( \frac{  (z+1)^{\frac{3}{2}}}{ \sqrt{| \zeta |}} SIN_k( \sqrt{| \zeta |}   \ln (z+1) ) \right) + C .   
\eeq
 For $k=0$, that is to be understood in the sense of  the limit $\zeta \rightarrow 0$, 
\[ m_0 = 5 \log  \left( (z+1)^{\frac{3}{2}}   \ln (z+1) ) \right) + C .     \]
%%%%%%%%%%%%%%
\\[0.5ex] \noindent
{\bf Angular size}
\\ \noindent
The great simplicity of Weyl universes results in simple  explicit expressions for  other
geometrico-physical properties.  For the Einstein-Weyl case, the angular size of objects with diameter $b$ in  distance $d$ is given by spherical trigonometry  for $k=1$,  because   the light cone structure is  conformally invariant  under deformation of the classical Einstein universe to the Einstein-Weyl one. Angular sizes of such objects  are   given by 
\beq \label{angular-size} \sin \frac{ \alpha  }{2 } = \frac{ \sin\frac{ b }{2 a } }{\sin \frac{ d }{ a} } ,\eeq
where $a:= \frac{H^{-1}}{ \sqrt{\zeta } }$ is the (ex-ante) radius of curvature of spatial sections.
%%%%%%%%%%%%%%%%%%
\subsection*{5.  Segal background and vacuum energy}
%\[0.5ex] \noindent
{\bf  Segal background}
\\ \noindent
I.E. Segal  proved the  mathematical existence of   a background equilibrium for the quantized Mawell field in the classical Einstein universe \cite{Segal:CMB1}. He proposed to consider this equilibrium radiation as an alternative explanation of the cosmic microwave background. 
   His existence result can be transferred to the Einstein-Weyl case.
 We should, however, {\em not  directly identify} the  CMB  with the Segal background, as Segal did. Because of the  long range directional properties of the CMB, observable among others by its anisotropy properties, it  appears impossible  to  interpret the CMB as a  direct expression of a background state of the Maxwell field in stochastic equilibrium. Any long range directional signals would have been absorbed by the stochastical equalization of the background.   If the CMB is due to the Segal equilibrium state, it can only  arise  as a redshifted {\em caustic phenomenon of the Segal background radiation emitted close to the observer's first conjugate point}   in the spatial fibre $S_{\kappa }$. 

\begin{figure}[t]
\center{\includegraphics*[height=5.0cm, width=9.0cm]{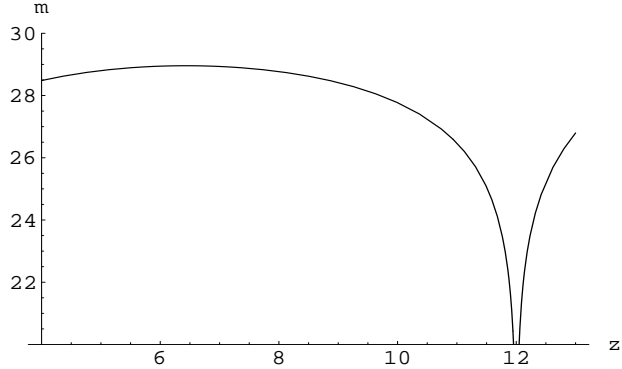}}
\caption{Relative magnitudes $m$ of  sources  at redshift $z$ (absolute magnitude  $M = - 19)$  and  refocussing of radiation in Einstein-Weyl universe  ($\zeta =1.5$)  close to the conjugate point, here  at  $z_{conj}\approx 12$  \label{fig refocussing}}
\end{figure}

The luminosity function $m_+ (z)$ for the Einstein-Weyl case reflects the  refocussing of the light cone at  the conjugate point by a rapid increase of luminosity  near  the corresponding redshift value, $z_{conj}$. Theoretically it even becomes singular  $m_+ (z) \rightarrow -\infty $ for $z \rightarrow z_{conj}$   (figure \ref{fig refocussing}). 
%%%%%%%%%%%%%%%%
\unitlength1cm
%%%%%%%%%%%%%%%

In this case, the Segal background  (which is of exact Planck characteristic) has to   lie in the infrared band and  can be considered as a kind of modified Olbers effect. Here the night sky is not as bright as Olbers expected, because its energy  is damped by the Hubble effect. It is not perceived directly, because in the  infrared band a separation of cosmic and other origins is very  difficult to achieve \cite{Dwek/Hauser}, and the background itself does not possess the properties of directed radiation. Only its caustic from the conjugate point does and has been redshifted by $z_{conj}$, the value of $z$ at the first conjugate point of the observer.

If the CMB is such a caustic phenomenon in an Einstein-Weyl universe with module $\zeta $,  the temperature of the Segal background can be calculated using (equ. (\ref{redshift Weyl universe})) at the first conjugate point with distance $d= \pi R$, $R = H^{-1} \zeta ^{-\frac{1}{2}}$:
\beq  \label{T-Segal}T_{Seg}  = (z_{conj} + 1) \, T_{CMB} = e^{\pi \zeta ^{-\frac{1}{2}}} T_{CMB} \; .  \eeq

The energy density  of a Planck radiation   of temperature $T$,
\[  \epsilon (T) =    \frac{\pi ^2}{15 ( \hbar c)^3} (k_B \,  T )^4  \; ,\quad \; \quad k_B \; \; \mbox{Boltzmann constant, } \;\; \] 
gives     the relative energy density of the  Segal background: 
\beq  \label{OmegaSegal} \Omega _{Seg}(\zeta )  = \frac{\epsilon (T_{Seg})}{\rho _{crit}} =    \frac{  \pi ^2 (k_B \,  e^{\pi \zeta ^{-\frac{1}{2}}} T_{CMB} )^4  }{15 ( \hbar c)^3 \rho _{crit}}   \eeq
$T_{CMB}$ and the Hubble constant $H= H_0$ have been  empirically determined with high  precision:
\beq \label{H_0 und T_CMB}
H_0 = 2.27 \, (\pm 0.2) \, 10^{-18} \; s^{-1}, \footnote{That corresponds to $ H_0 \approx  70\, \; km s^{-1} Mpc^{-1} \, (\pm 8 \%)$.} \quad
T_{CMB} = 2.725 \, (\pm 0.004)° K \;.
\eeq
Equation  (\ref{OmegaSegal})  thus establishes a functional relationship between $\zeta $ and $\Omega_{Seg} $.
%%%%%%%%%%%%%%
%%%%%%%%%%%%%%
\\[0.5ex] \noindent
{\bf Energy densities}
\\ \noindent
For the total r.h.s tensor $T$ of the Einstein  equation we now have to take  the  radiation
 contribution $T_{rad} = T_{Seg}$ of the Segal background to $T$ into account,
\beq T = T_m + T_{Seg} + T_{\Lambda  }  \; . \eeq
The energy density and pressure equations, (\ref{energy density condition}),  (\ref{pressure condition}),
and  (\ref{OmegaSegal}) give  3 independent conditions  for the   geometrical and matter parameters $\zeta , \Omega _m$, $\Omega _{Seg}$, $ \Omega _{\Lambda  }$.   A
 single additional empirical datum, like the mass density parameter $\Omega _m$,  suffices to specify the model completely. Although $\Omega _m$  is  much less precisely determined empirically, the present interval 
\beq \label{Omega_m interval} 0.1 \leq \Omega _m \leq 0.4 \eeq
 gives a comparably sharp constraint for the module, $ \zeta \approx 1.58\; \pm 12 \% $, because $\Omega _m$ is small in comparison with $ \Omega _{\Lambda  } + \Omega _{Seg}$ (see table 1). 

\begin{table}[h]\centerline{ \bf  Table 1: Balanced density parameters depending on  $\Omega _m$ } \vspace{01ex} 
\begin{center}
\begin{tabular}{|| l || c| c | c | c ||}  \hline
$\Omega _m$& 0.1 & 0.2 & 0.3 & 0.4  \\  \hline \hline
$\Omega _{\Lambda }$ & 1.25 & 1.23 & 1.22 & 1.20 \\  \hline
$\Omega _{Seg}$ & 1.20 & 1.13 & 1.07 & 1.00 \\  \hline
$\zeta $ & 1.55 & 1.57 & 1.59 & 1.61 \\ \hline 
\end{tabular} \par \end{center}
{ \small $\Omega _m$ matter density parameter,   $\Omega _{\Lambda }$ vacuum energy density, $\Omega _{Seg}$  radiation energy density of the Segal background,  $\zeta $ geometrical module of Einstein-Weyl universe  }
\end{table}
The error interval  for  $\Omega_{\Lambda } $ and $\Omega _{Seg}$   is $\pm 8 \%$, mainly due to the error of $\rho _{crit}$ (inherited from $H_0$). The error bound for $\zeta $ is the interval of table 1 plus $8 \%$. We thus arrive at the following values of the parameters for an  Einstein-Weyl geometry of  cosmic space-time:
\beq   \label{parameters E-W model}  \Omega _m \approx 0.25  \;\;   (\pm 60 \%), \; \Omega _{\Lambda  }\approx 1.23, \; \Omega _{Seg} \approx  1.1 \;\;   (\pm 8 \%), \quad   \zeta \approx 1.58\; (\pm 12 \%),  \eeq

Here we have presupposed that the energy and pressure conditions hold. In the next paragraph we discuss how such an assumption may be physically founded. 
%%The energy densities satisfy  the condition  
%%\beq   \label{tuning}  \Omega _m + 2 \, \Omega _{Seg}  = 2  \Omega_{\Omega }  \; .  \eeq
 %%%%%%%%%%%%%%
%\\[0.5ex] \noindent
\newpage \noindent
{\bf Stability questions}
\\ \noindent
The precise understanding achieved during the 1960s for  necessary and sufficient conditions for the appearance of singularities  and their structural properties   contributed strongly  to discredit not only the classical static solutions but also all other cosmological solutions without initial singularity \cite{Raychaudury,Penrose:Singularities,Hawking/Penrose,Hawking/Ellis}. The  deep existence theorems for these types of  singularities  depend essentially on the decision of classical relativity for the semi-Riemannian structure and its   inbuilt decision for  Riemann gauge as matter gauge in the sense of our postulate \ref{clocks}. 

Clearly Weyl universes do not contradict these theorems (they  have an obvious  ``initial singularity'' in Riemann gauge), but the singularities loose the immediate geometrical and physical importance  of the classical geometric setting.  The upper bounds for geodesic lengths in the  estimations of the theorems no longer prove  incompleteness in other gauges. Our simple examples, the Weyl universes, are  incomplete in Riemann gauge but  geodesically complete in Hubble gauge. 

Geometry  alone (in the sense of IWG)  does not tell anything about the ``real'' existence of an initial singularity and/or stability. Turning the table we consider the possibility that  quantum vacuum processes, expressed summarily by $\Omega _{\Lambda }$, the background equilibrium state of the quantized Maxwell field $\Omega _{Seg}$ and the matter component $\Omega _m$ of the energy content in the universe stand in a balance expressed by
\beq \label{balance equation} \Omega _{\Lambda } =   \Omega _{Seg} + \frac{1}{2} \Omega _m \; .\eeq
Such a balance cannot be considered a  result of the physical geometry of GRT. If it corresponds to physical reality, it is  due  to an {\em exchange equilibrium} in large means between the cosmic vacuum, the Maxwell field and the matter content of the universe.

Equation (\ref{balance equation}) implies the hyle condition $p = - \frac{\rho }{3}$. The equation of motion for the deformation of Weyl universes (\ref{equation of motion Weyl universes}) tells us that in this case $f''=0$ and $f'= const$. This does not lead outside the class of Weyl universes. Without loss of generality we may assume the deformed model to be Hubble gauged, i.e., $f'=0$. In this sense Weyl universes are stable.\footnote{Surely this argument can be refined. For the special type of the  Milne universe that  has already been done. Compare \cite[27]{Rendall:Existence}.} 

Thus the use of Weyl universes does not presuppose  to  assume an  a-priori static geometry. Time homogeneity of the Weyl metric, both of its Riemannian component and of its length connection, may be considered  as the geometrical result of a  dynamical exchange  balance between the quantum vacuum, the Maxwell field (Segal background), and matter proper. In this sense {\em statics becomes a special case of dynamics}  on the level of cosmology like in other contexts. 

The new approach is no longer static in the sense of the 19th century or the first generation of relativistic cosmologists of the 20th century, Einstein, de Sitter and others. To make this difference clear we  better call the new models ``neostatic''. They only appear to be static in large means; in smaller regions warps and fluctuations modify the picture and may even dominate.
%%%%%%%%%%%%%%
\\[0.5ex] \noindent
{\bf Vacuum energy}
\\ \noindent
In recent  background field (BF) studies of the quantum vacuum with Chern-Simons actions, C. Castro has  calculated a geometric mean of vacuum energy density  by   assuming    the existence of an infimum  $l$ and a supremum  $L$  of physically meaningful length scale  (lower and an upper ``cutoff'') \cite{Castro:ADSetc,Castro:Clifford_Spaces}: 
\[ \rho _{vac} = [\hbar c] (l \, L)^{-2} \; \]

The infimum is naturally given by the Planck length $l_{Pl} = c^{-1}(\frac{\hbar N}{c})^{\frac{1}{2}}$. With   the Planck energy $E_{Pl}= c^2 (\frac{\hbar c}{N})^{\frac{1}{2}}$, Castro's vacuum energy density becomes
\beq  \label{Castros vac energy}         \rho _{vac} =  E_{Pl} l_{Pl}^{-1} L^{-2}  = c^4 N^{-1} L^{-2}              \eeq
Adopting the Hubble length $H_1^{-1} = c H_0^{-1}$  as  (infrared) cutoff scale, which seems plausible although slightly {\em  ad-hoc} in the frame of {\em standard cosmology}, Castro observed that his result ``reproduces precisely the {\em observed} value of the vacuum energy density'' \cite[974, 1022]{Castro:Clifford_Spaces}.\footnote{The 
ad-hoc value  $L=H_1^{-1}$ leads to $\Omega _{vac } = \frac{8}{3 \pi} \approx 0.85 \sim 0.75 $. That is a great achievement in comparison to the 120 orders of magnitude error of the quantum fluctuation calculation. }

Castro's result becomes  more convincing in the context of Einstein-Weyl models. There  a {\em natural supremum} for length scales is given by the distance  $L = \pi r$ to the next conjugate point (the semi-circum\-ference of the spatial 3-sphere). With $L^{-2}= \pi^{-2} \kappa =  \pi^{-2}  H^2 \zeta $, the relative value of Castro's vacuum density  is  
\beq \label{Omega_vac Castro} \Omega _{vac} = \frac{  \rho _{vac}}{  \rho _{crit}}  = \frac{8}{3 \pi} \zeta \; .\eeq
For $\zeta = 1.55$ Castro's calculation gives $\Omega _{vac} \approx  1.32$, only $6 \%$ above the value for   $\Omega _{\Lambda }$ at $\zeta =1.55$,   $\Omega _{\Lambda } \approx 1.25$ (table 1). 
  This is in fact a  striking {\em agreement  inside the error interval} $ 8 \%$.  For higher values of $\zeta $ the error increases moderately. It leaves the error interval at $\zeta =1.58$.

In the Weyl geometric approach,  the vacuum energy density is  constant in large means. {\em It is not affected by the dynamical dark energy anomaly of the standard approach.} If C. Castro's calculation can be justified in a working quantum field theoretic frame (e.g.,  the quantized Clifford group unification proposed in  \cite{Castro:Clifford_Spaces}), we have a satisfying solution of the vacuum constant riddle.

Our equations (\ref{energy density condition}),   (\ref{balance equation}) and (\ref{Omega_vac Castro}) indicate how a (fictitious) model universe {\em without} ponderable mass, might look like:
\beq \label{ether universe} \Omega _{\Lambda } = \Omega _{Seg} = \frac{\zeta +1}{2}=  \frac{8}{3 \pi} \zeta\; , \quad \quad  \Omega _m = 0 \eeq
It characterizes a pure ``ether'' state of the universe, in which the quantum vacuum excites the electromagnetic field sufficiently to induce a Segal background, such  that a time-homogeneous geometrical state is acquired. In the case of (\ref{ether universe}) the negative and positive pressures of both components supplement each other such that the hyle equation (\ref{hyle}) is satisfied.

If we join equations (\ref{OmegaSegal})  and (\ref{H_0 und T_CMB}), the geometrical module of a matter free model universe with a Castro vacuum and Segal background is determined as $\zeta \approx 1.43$. Then  we find $\Omega _{\Lambda } = \Omega _{Seg} \approx 1.22$. These values characterize the Einstein-Weyl geometry of a  background cosmic gravitational  field induced by the cosmic vacuum  and a counterbalancing Segal radiation. Our observed universe  appears as a modification of this background system by   a small amount of ponderable matter (cf. equ.
 (\ref{parameters E-W model})). Matter breaks the identity of $ \Omega _{\Lambda }$ and $\Omega _{Seg} $ and deforms the balancing condition slightly. 

\subsection*{6. Predictions,   observational constraints and perspectives}
\begin{figure}[h]
\center{\includegraphics*[height=5.0cm, width=9.0cm]{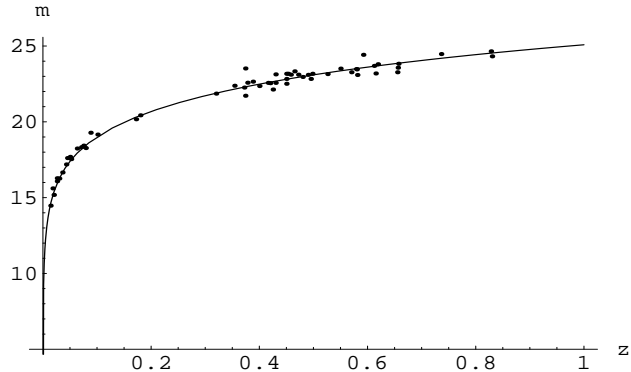}}
\caption{Relative magnitude $m$ of supernovae Ia, $M= - 19.3$, at redshift $z$ in Einstein-Weyl model with $\zeta \approx 1.5$ (line). SN Ia data (dots) from (Perlmutter e.a.1999). \label{fig SNIa}}
\end{figure}

Finally we have to discuss whether Einstein-Weyl universes have an empirical ``surplus
value'', i.e., whether they have   empirically testable consequences beyond those  used to determine the model parameters.\footnote{Compare also \cite{Scholz:Extended_Frame} and the semi-popular exposition \cite{Scholz:BerlinEnglish}.}
In this sense we now discuss  some  predictions  relative to the observational input data, equ. (\ref{H_0 und T_CMB}) and $ 0.1 \leq  \Omega _m \leq 0.4$.
%%%%%%%%%%%%%%
\\[0.5ex] \noindent
%\newpage  \noindent
{\bf Supernovae Ia}
\\ \noindent
Once the input data for $H_0, T_{CMB}$ and $\Omega _m$ have been fixed (equs. (\ref{H_0 und T_CMB}), (\ref{Omega_m interval})), the {\em  luminosity redshift data of supernovae  Ia } \cite{Perlmutter:SNIa} are  {\em predicted} by our model (figure \ref{fig SNIa}). The  fit quality compares well with the one achieved by the Friedmann-Lema\^{\i}tre model class.\footnote{In the F-L model class the best fit of the data set  of \cite{Perlmutter:SNIa} has  dispersion  $1.21 \sigma _{dat}$ ($ \sigma _{dat}$ = mean square error of the observational data).  The best fit in the Einstein-Weyl model class is below  $1.23 \sigma _{dat}$.} Although the best fit  for   Einstein-Weyl  models is reached at    $\zeta \approx 1.8$,  the  fit quality decreases only insignificantly with changes of the module in the interval $1.5 \leq \zeta  \leq 1.61$.\footnote{Rise from $1.227 \sigma _{dat}$ for the best fit to $1.228 \sigma _{dat}$, still below $1.23 \sigma _{dat}$. Cf. \cite{Scholz:Extended_Frame}.} 
%%%%%%%%%%%%%%%%
\unitlength1cm
%%%%%%%%%%%%%%%

Since the late 1990s the supernovae data have   widely been considered as a ``proof'' for  the correctness of the present standard model and as  clear empirical evidence for an accelerated expansion of the universe. Physicists started to be convinced that an unknown type of ``dynamical'' dark energy exists and blows our universe apart. The comparably good fit in the Weyl universe class   relativizes such  straight forward realistic interpretations of Friedmann-Lema\^{\i}tre  models. They provide a  sound, general relativistic, alternative interpretation in which the supernovae luminosities are predicted without such strange  model properties.
%%%%%%%%%%%%%%
\\[0.5ex] \noindent
{\bf Anisotropies of  cosmic microwave background}
\\ \noindent 
For  $\zeta \approx  1.5$ the contributions  to the  {\em anisotropy signal} between $0.8 °$ and $ 1 °$  resulting from 
   regions with small redshift, $z<1$, are due to inhomogeneities  on cluster level. A strong correlation between anisotropies and cluster positions for small redshifts has  been empirically observed. It is  discussed   in \cite{Shanks:WMAP}.  The authors  consider the corresponding temperature loss to be caused by the Sunyaev-Zeldovich effect. For larger redshifts,  $1 \leq z \leq 6$,  inhomogeneities on the size level of superclusters come into the play. The observed temperature losses are here caused by  crossings of regions in which the Hubble effect (due to the gravitational field or  exchange with the Segal and/or vacuum background) may be above average.     The Einstein-Weyl model lets us  expect   temperature fluctuations in correlation with  the difference of the number of clusters and super clusters passed by a geodesic path to the next conjugate point   from the statistical mean. A peak is expected at angles $\approx 0.9$, corresponding to a angular momentum $ l \approx  200$. It is caused by super clusters in a wide band about the equatorial 2-sphere of the spatial fibre. The interval $1 \leq z \leq 6$ covers about $80 \%$ of the volume of the spatial 3-sphere. These effects dominate the total anisotropy signal. The distribution of objects at  different size levels leads to typical fluctuation of the overall  stochastical distribution in the anisotropy signal. The decomposition  into  higher momenta of spherical harmonics has still to be analyzed. 
%%%%%%%%%%%%%%
\newpage \noindent
%\\[0.5ex] \noindent
{\bf Quasar frequencies}
\\ \noindent
As a bonus, unknown to the standard approch,  the Einstein-Weyl model allows to {\em predict  the relative distribution of quasars} and other cosmic objects {\em over redshift} by the geometry of the light cone.  The volume of light cone layers corresponding to equal increments in redshift  increases obviously until the passage of the equatorial sphere  and starts to decrease thereafter.

 A quantitative evaluation (figure 3)  shows that the volume increments behave very much  like the observed {\em quasar frequencies} reported in \cite{SDSS:2003b}. From the perspective of Einstein-Weyl models, quasars appear to be approximately  equally distributed on  large cosmic scales, at least until the frequency peak at $z \approx 2$. 
For larger $z$  the incremental volumes decrease, but not as fast as the observed quasar frequencies.
Apparently selection effects (most importantly a rapid decrease of sensibility of CCD detectors in the frequency range of quasar light emission above $z\approx 2$) take over here.
%%%%%%%%%%%%%%%%
\unitlength1cm
\begin{figure}[h]
\center{\includegraphics*[height=5.0cm, width=9.0cm]{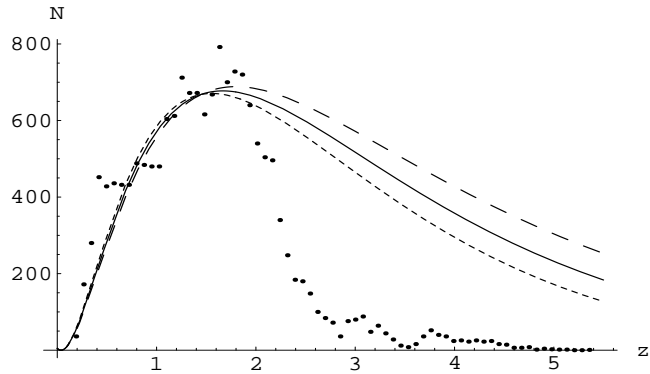}}
\caption{Number of  equidistributed objects observable in equal  redshift intervals of length $\Delta z =0.076 $ in Einstein-Weyl universes,  $\zeta =  1.3$ (coarse dashing), $\zeta =  1.5$ (undashed)  and $\zeta = 1.7$ (fine  dashing), normed by a total number $N=14610$ up to $z\leq2.28$. Comparison with  quasar counts of SDSS,  first data release (dots). \label{fig quasars} }
\end{figure}

 A similar increase and peak close to $z \approx 2$ (not yet the flank of the decrease)  has been observed for $\gamma $-ray bursts \cite{Lamb:GRB}. 
In the Einstein-Weyl models these formerly unrelated observational regularities appear as simple results  of the light cone geometry.
%%%%%%%%%%%%%%
\\[0.5ex] \noindent
{\bf A less critical frame  for structure formation}
\\ \noindent
If the Einstein-Weyl model is a physically adequate representation of cosmic time-space geometry in the large, the idea of a {\em global} evolution of the cosmos turns out to be a fiction. Evolution, transformation  and disintegration  of structures then make sense only on a ``regional'' level. In this framework there is no reason to expect an increase of the mean metallicity of galaxies over very long (``cosmic'') time intervals. Structure formation becomes a much more open field. It is no longer subject to to the precarious `balance' between  the ``expansion dynamics'' of the universe and attractive forces of exotic cold dark matter. 

The time perspective is opened, and quasars need no longer  be considered as products of the formation period of galaxies.  In the Weyl geometric frame  it appears at least as natural to consider quasars as  products of accreted mass in nuclei of {\em old} galaxies. In the neighbourhood of quasars and active galactic nuclei there seem to arise very high temperatures. Perhaps they provide  distributed loci for very high temperature contributions  to nucleosnythesis which have to exist in addition to stellar nucleosynthesis and are looked for shortly after the big bang in the standard approach.

Of course, these questions can only  be answered in a further  interplay of empirical studies and theoretical investigations. 

%\\ \noindent
\subsection*{7.  Summary and conclusions}
We have seen that only small changes of Dirac's version of integrable Weyl geometry are necessary to turn it into  a conceptual frame for  a physically meaningful  conservative extension of GRT. It  makes an otherwise suppressed {\em scale invariance of the Einstein equation}  visible. The standard explanation of  cosmological redshift by space expansion and a ``more physical one'' (Weyl) by an  energy loss of photons can now be treated  on the same mathematical footing. In this sense, IWG becomes an excellent conceptual frame for an {\em extension of  Einstein's equivalence principle}. 

Dynamical consequences of the  modification of GRT  inside the solar system are several orders of magnitude below observational errors, while they do matter on the size level of clusters and above. That has consequences for the calculation of dynamical mass density $\Omega _m$ from observational data. The present method for calculating dynamical mass  turned out to be {\em inconsistent with the framework of   Friedmann-Lema\^{\i}tre cosmology, while it is consistent with IWG (Hubble gauge)}.  

Our investigation of the simplest conceivable Weyl geometric models, the Weyl universes, demonstrates that many cosmological phenomena can well  be represented in a a time homogenoeus (``neo-static'')  time-space geometry with  a single geometrical parameter.  Those with positive 
ex-ante spatial curvature, the Einstein-Weyl universes, seem to be promising candidates for cosmic geometry  and avoid  the anomalies of the standard approach. They allow to predict the supernovae luminosities as well as the standard approach, give an explanation of the cosmic microwave background and its anisotropies,  different from the standard one, and are consistent with dynamically determined matter densities. Moreover, they give a basically geometric explanation for the characteristic behaviour of quasar frequencies (or those of $\gamma $-ray bursts) in dependence of redshift. 

The assumption of a Segal  equilibrium as background of the electromagnetic field even makes  our models appear quite realistic. At  the side of the $\Lambda  CDM$ ($\Lambda  $, cold dark matter)  model  of the standard approach with $(\Omega _m, \Omega _{\Lambda  }) \approx (0.25, 0.75)$ we can now place an  IWG model  consistent with the present estimations of mass density and other data. It 
assumes only {\em ordinary} dark matter (mostly molecular hydrogen) and a {\em vacuum energy density} withouth the irritating dynamical behaviour of the standard approach. 

The model assumes a Segal background, as an equilibrium state of classical radiation. It may be abbreviated by  $\Lambda  SDM$ ($\Lambda  $, Segal background, ordinary dark matter). A typical specification in terms of today's accepted mass densities  is $(\Omega _m,\Omega _{\Lambda  }, \Omega _{Seg}) \approx (0.25,1.23 , 1.1)$ with a geometrical module $\zeta \approx 1.58$. These data lead to an excellent agreement   with C.Castro's calculation of the vacuum energy density from the assumption of an infimum of  physically valid lengths  at the Planck scale  $l_{Pl}$ and a supremum at the distance to the antipodal point on the spatial sphere. Other  mass densities of the  presently accepted interval do not destroy the consistency with the other empirical data which are considered here, although Castro's vacuum energy density moves beyond the bounds of the error interval for $\Omega _{\Lambda }$. 

Our discussion of observational constraints shows  that it is not necessary to project the complexity of cosmology  onto the physico-geometrical framework. While for the characterization of  the matter and energy content of ``the world'', an  introduction of more  parameters is  clearly appropriate, geometry may stay comparably simple if a proper conceptual perspective is chosen.

We  see that the direct  translation of the empirically determined parameters characterizing matter and energy distribution into properties of an imputed space-time expansion, which is  characteristic for the standard approach, is {\em no necessary consequence of  general relativity}, once it is understood in the extended frame of IWG.   It  is a {\em  result of the decision for Riemann gauge}, inbuilt in the standard theory. We have found first evidences  that this decision may rather be a fault than a strength of the present standard model of cosmology.  Surprising consequences of this strategic theory decision are usually not  considered as  primarily mathematical model features which arise as unhappy consequences from fitting the phenomena. They are usually thought to  represent some transcendent temporal physical ``reality'',  at least in the official 
self-image of  mainstream cosmology.\footnote{A different opinion is expressed by the signers of the letter \cite{OpenLetterCosmology}. } Although such a basic belief cannot  be disproved by alternative theories, the Weyl geometric approach shows   that it is  not necessary  to share it. A  general relativistic symbolic representation of the universe in the large is possible along different lines.
In particular the discussion about  ``dynamical dark energy''  seems to contain more  social imaginery, than physicists  usually like to admit. We now see that it is {\em no  necessary} outcome of theory development and of empirical findings. 

All in all, the Weyl geometric approach to general relativity opens a path towards  analysing the open questions of cosmology deeper than before and with less inbuilt conceptual restrictions than in the classical 
semi-Riemannian paradigm. This will remain so, even if it should turn out that Weyl universes offer just another over-simplified geometrical frame for   the  macro-cosmos we consider ourselves part of.
%%%%%%%%
%\newpage \noindent
\vspace{20mm}
\section*{Appendices}
%%%%%%%%%%%%%%
%\\[0.5ex] \noindent
\subsection*{Appendix I: Basics of integrable Weyl geometry}

\noindent {\bf  Weylian metric} \\ \noindent
We work with a  Weylian metric on a 4-dimensional differentiable
manifold $M$, given  by a Lorentzian metric $g$ of signature  $(-,+,+,+)$   and a real valued differential 1-form 
$\varphi$.\footnote{\cite{Weyl:InfGeo,Eddington:Relativity,Bergmann:Relativity,Folland:WeylMfs,Ehlers/Pirani/Schild,%
Dirac:1973,Canuto_ea,%
Maeder:WeylGeometry,Hehl_ea:Kiel_II,Tiwari:Weyl_geometry,%
Drechsler/Tann,Frauendiener:ConfInf,%
Varadarajan:Connections,Scholz:Extended_Frame}}
 After choice of coordinates $x=(x^0, x^1,
x^2, x^3)$, the metric is given locally by
\beqa  g & =& (g_{ij}), \; \; \, ds^2 = \sum_0^4 g_{ij} dx^i dx^j, \\
\varphi & =& (\varphi _i) , \; \; \; \, \varphi = \sum \varphi_i dx^ i  .\eeqa
 $g$ is  the {\em Riemannian component} of the Weylian
metric and $\varphi$ its {\em length} or {\em scale connection}.  The pair $(g, \varphi)$ defines  a
{\em gauge} of the metric. It can be changed to another one, $(\tilde{g}, \tilde{\varphi}) $,  by a  {\em gauge transformation}  
\beq \tilde{g}(x) = \Omega^2 (x)\, g(x)= e^{2 \Phi (x)} g(x) \, , \;\; \tilde{\varphi} =
\varphi - d \Phi \, ,
 \eeq
with a real valued function $\Phi$ on a local neighbourhood. 

The  structure of a {\em  Weylian
manifold} on  $M$ is given (locally) by an  equivalence class $[g,
\varphi]$ of gauges. It is important  to keep in mind, that the Weylian metric is {\em not changed} under a gauge transformation; only its representation by Riemannian component and length transfer is. This is comparable to the change of representations of a semi-Riemannian metric under coordinate transformations. For a full characterization of the change of  physical  reference systems  we   have to take gauge covariance  into account,  in addition to  coordinate covariance. 

{\em Warning}:  In his fist presentation of the newly developed gauge geometry to a physical audience, Weyl used a different sign convention  for the gauge transformation of the differential form \cite{Weyl:EFT}: 
\[ \tilde{\varphi} =
\varphi + d\, log \Omega   \]
That did not affect the field theoretical intentions of this article and made his gauge connection look more closely like a potential as used in the electromagnetic literature. It {\em affected}, however, the possibility of the straight forward calculation of a {\em  gauge invariant length transfer} crucial for Weyl's geometrical approach, see equ. (\ref{length comparison}) and (\ref{calibration-transfer}) below. The  length transfer  function had now to be calculated by the reciprocal of the transfer integral. This geometrically anti-intuitive sign convention was introduced into the physical literature by W. Pauli's publications \cite{Pauli:WeylsTheorie,Pauli:EMW} and  \cite{Eddington:Relativity}. Weyl never again used this geometrically maladroit  sign change convention.\footnote{Cf. \cite{Weyl:Erweiterung,Weyl:eGRT}, 3rd and later editions of \cite{Weyl:RZM}.}
Unfortunately it has become  standard in the physics literature and may have contributed to the obstacles for acquiring a proper geometrical understanding of Dirac's retake of integrable Weyl geometry in the 1970s.
Weyl's original sign choice agrees with the definition of  gauge transformation in modern differential geometry and will be used here. 
%%%%%%%%%%%%%%%%
\\[0.5ex] \noindent 
{\bf  Length transfer and scale covariant fields} \\ \noindent
Any scalar, vectorial or tensorial quantity $w$ on $M$
which transforms under rescaling of the metric by $\Omega (x)^2  = e^{2
\Phi (x)}$ like
\beq  \tilde{w}(x) = \Omega(x)^k w(x), \; \; k \in \Z\eeq
will be called a {\em scale covariant field} of {\em weight} 
$k$. More precisely, a scale covariant field is an equivalence class of scalar, vector or tensor valued functions on $M$ with representative selected by a choice of  gauge.\footnote{More formally in  \cite{Scholz:Extended_Frame}; this characterization is due to M. Kreck.} Dirac took up   Eddington's terminology of \cite{Eddington:Relativity}. Thus in his papers
\cite{Dirac:1973,Dirac:LNH} and the literature following him,
scale covariant fields are called  {\em co-covariant} scalars, vector
or tensor fields.  

 We write $k = [[w]]$ for the gauge weight of $w$. The Riemannian component of the metric is of
weight $[[g]] =2$, $g^{ij}$ is of weight $-2$ etc., while the length connection
is no  scale covariant field at all. It transforms like a connection in the modern
sense of differential geometry and is, in this sense, a gauge quantity
{\em sui generis}. 

If the length connection is integrated along a differentiable path
$\gamma (s)$,  $s \in I = [0,1]$(or  another  interval $I$), the integral can be used as a {\em  length transfer} function $l(p_0, p_1)$  between
the endpoints $p_0:= \gamma (0)$ and $p_1:= \gamma (1)$
\beq \label{length-transfer}  l(p_0,p_1):= e^{\int_{0}^{1} \varphi (\gamma ')
ds  } \, . \eeq 
Comparing vectors at different points $p_0, p_1$, say $\xi \in T_{p_0}M$ and $\eta \in T_{p_1}M$, by the quotient
\beq \label{length comparison} \frac{|\eta |_g}{ l(p_0, p_1)|\xi |_g}   \eeq
gives a {\em gauge independent} criterion which is in general  path dependent (although not so for integrable Weyl geometry).

The concept of length transfer generalizes naturally  to scalar scale covariant fields of any gauge weight. 
If $l(\gamma (s))$   is the length transfer with respect to a gauge $(g, \varphi)$ and a curve $\gamma $ in M, a scalar scale covariant field  $f$ of gauge weight $k$, defined along $\gamma $, is said to  propagate by {\em calibration transfer} along $\gamma $, if
\beq  \label{calibration-transfer} f(\gamma (t)) = l(\gamma (t))^k f(\gamma (0)) . \eeq

The length connection is called {\em integrable}, if such integrals
between the same points are path independent.  On the local level that
is equivalent to
\[ d \varphi = 0 \, .\] Then $\Phi(x) = \int \varphi(\gamma ')$, and the Weylian metric can be given a (semi) Riemannian form by rescaling to $\tilde{g} :=\Omega^2 g $,
with  $\Omega (x) := l(p_0,x)$. This leads to the  {\em Riemann gauge}
$(\tilde{g}, 0)$ with vanishing length connection.
Here we consider only {\em integrable}  Weylian
metrics. Compatibility with quantum physical structures makes
this restriction necessary  \cite{Audretsch_ea}. 
%%%%%%%%%%%%%%%%%
\\[0.5ex] \noindent 
{\bf Affine connection} \\ \noindent
Every Weylian metric (also a  non-integrable one) specifies a
uniquely determined  compatible {\em affine connection} $\Gamma $. Its
components  differ from the 
Levi-Civita connection $_g\Gamma $ of the Riemannian component $g$ of
the Weylian metric in a chosen gauge $(g, \varphi)$ by   
\beq \label{Christoffel}  \Gamma^i_{jk} =   {}_g\Gamma^i_{j k} + \delta ^i_j \varphi _k +
\delta ^i_k \varphi _j - g_{jk} \varphi^i \, . \eeq
Here $ \delta ^i_j $ denote the Kronecker delta ($ \delta = diag (1,1,1,1)$).
$\Gamma$ is an invariantly defined geometrical object,
although  the right hand side contains  a gauge dependent expression of
it. It leads to a {\em gauge invariantly defined 
 covariant differentiation} $\nabla _{\Gamma}$ of vector and tensor
fields on $M$,
 in the sense of Riemannian differential geometry, and a {\em gauge invariantly defined geodesic} $\gamma$, $\gamma '(s) = u(s)$,
\beq \label{invariant_geodesic}  \frac{d}{ds }u^i +(\nabla_{\Gamma})^i_{j} u^j  = \frac{d}{ds
}u^i +_g\Gamma^i_{jk}u^k u^j + 2 \varphi_j u^i u ^j - g_{jk}u^j u^k \varphi^i = 0 \, . \eeq

The Riemann  curvature tensor $R = (R^i_{jkl})$ is gauge invariantly defined by the Weyl-Levi-Civita connection $\Gamma$, and so is the Ricci curvature $Ric = (R_{ij}), \; R_{ij} = R^k_{ikj}$ which is directly derived from it. Only the scalar curvature  $\bar{R}= g^{ij}R_{ji}$  depends on an  additional use of the Riemannian component of the Weylian metric; it is a scalar scale covariant field of weight $[[\bar{R}]] = [[g^{ij} ]]= - 2$. 
%%%%%%%%%%%%%
\\[0.5ex] \noindent 
{\bf  Gauge covariant differentiation} \\ \noindent
  In order to give better expression to physical properties in Weyl geometry Dirac introduced the concept of a {\em gauge (scale) covariant} differentiation $D_{\Gamma}$ of scale covariant fields in addition to the gauge invariant differentiation $\nabla _{\Gamma}$. If $w = (w^i)$
is a Weyl  vector field of weight $[[W]] = m$, its scale covariant
differential is given by 
\beq D_{\Gamma} w := \nabla_{\Gamma} w  + m
\varphi \otimes w \,,\eeq
 in coordinates
\[  ( D_{\Gamma} w)^i_j := \partial _j w^i + \Gamma^i_{jk} w ^k + m
\varphi_j w^i  \, , \]
with the abbreviation
$\partial_j := \frac{ \partial }{\partial x^j }$.
Accordingly, a scalar scale covariant field $a$ of weight $[[a]]= m$, has  gauge covariant differentiation
$(D_{\Gamma} a)_j = \partial_j a +m \varphi_j $.
For tensors of higher order, the length connection terms of the gauge
covariant differentiation are formed analogously to the terms in the
affine connection (see 
\cite[appendix A]{Canuto_ea}). Here we will use only the gauge covariant
differentiation of vector and scalar scale covariant fields. We will abbreviate the
notation in the sequel by writing
{\[   D:= D_{\Gamma} \, ,\]
where ambiguities can be excluded by the context.  

Dirac's gauge covariant differentiation  enhances  the mathematical language of Weyl geometry and adapts it better to  physical purposes. For
example, the definition of  a  geodesic by Weyl's covariant
differentiation $\nabla_{\Gamma}$ is  gauge invariant.
 If $\gamma $,
with tangent field $u(s) := \gamma '(s)$, is a gauge invariant
geodesic, its norm is a scalar scale covariant field of weight 1, because   $|u|^2 =
g(u,u)$ and $[[g(u,u)]] = [[g]] + [[u]] + [[u]]= 2 + 0 + 0 = 2$. Therefore a gauge invariantly defined
geodesic  necessarily ``changes length'' under gauge
transformations not only in finite intervals but also infinitesimally, because its tangent field   $u$  is a
well defined classical vector field along the curve. For the expression
of general relativistic particle mechanics in terms of Weyl geometry
this is a disturbing behaviour. Geodesics have to be rescaled before
they can be used to describe particle paths and energy-momentum of
particles (including photons). That was not only critical  for Weyl's original proposal of a unified field theory; it also would lead to unnecessary complications for the weak extension of general relativity we use here.\footnote{Compare the clumsy  treatment of
cosmological redshift in 
\cite{Scholz:Extended_Frame}, which can be considerably simplified by the use of
Dirac's geodesics. See below. }
%%%%
\\[0.5ex] \noindent 
{\bf  Gauge covariant geodesics, distance} \\ \noindent
Thus the  expression of general relativistic structures in terms of Weyl
geometry is greatly facilited by Dirac's definition of a {\em scale
covariant geodesic} $\gamma(s)$,  $u :=\gamma ' $,
of weight $[[u]]=-1$. It arises from an invariant (Weylian) geodesic by scale covariant reparametrization (weight -1) and can be defined
 by the differential equation
\beq \label{def-geodesic}
  \frac{d}{ds}u^i   +  D^i_j u^j   = \frac{d}{ds
}u^i +\Gamma^i_{jk}u^k u^j - \varphi_j u^i u ^j  = 0 .\eeq
Introducing the  term  $_g \Gamma$ of the Riemannian component that is
\[  \frac{d}{ds
}u^i +_g\Gamma^i_{jk}u^j u^k +  \varphi_j u^i u ^j - g_{jk}u^j u^k \varphi^i = 0 \, . \]
Formally, the crucial difference to (equ. (\ref{invariant_geodesic})) lies in  the different factor of the $\varphi_j$ term resulting from the weight correction term $-\varphi_j u^j u^j $ from scale covariant differentation.

The parametrization of a scale covariant geodesic is gauge
dependent by definition; its tangent field $u$ is of weight $-1$ (like it
should be for the symbolical expression of energy-momentum of a
particle). Because of 
\[ [[g(u,u)]] = [[g]] + 2 [[u]] = 2 - 2 = 0 \,  \]
the norm $|u|$ is  of weight $0$, i.e., it is a classical scalar (an
``in-scalar'' in Eddington's and Dirac's slightly idiosyncratic terminology). In  semi-Riemannian gauge, the norm
is a constant, e.g. $|u|^2=-1$, for a timelike curve.  As this is true 
in any gauge,    gauge covariant geodesics are well adapted to
express  general relativistic particle behaviour in the Weyl geometric
framework.
 We have therefore  adopted Dirac's {\em scale covariant definition of geodesics}
 in the sequel (principle (\ref{trajectories}) below).  To ensure consistence with lightlike geodesics, we also define ``scale covariant'' nullgeodesics by equ. (\ref{def-geodesic}), although the attribute is no longer to be understood literally. 

{\em Distances} will be measured in the Riemannian component of each gauge  $(g, \varphi)$  along  geodesics $\gamma$, $\gamma '= u,$  (gauge covariant or not): 
\beq d_{(g, \varphi)}(p_0, p_1 ) := \int_{\tau _0}^{\tau _1} \sqrt{|g(u(\tau ), u(\tau ))|} d \tau \quad \mbox{for } \; 
p_0 = \gamma (\tau _0), \, p_1 = \gamma (\tau _1)  \;  \eeq
By definition they are gauge dependent. Where context makes clear which gauge is referred to, we simply write 
$ d_{(g, \varphi)}(p_0, p_1 ) = d(p_0, p_1 ) $.
%%%%%%%%%%%%%%
\\[1ex] \noindent
\subsection*{Appendix II: Low velocity orbits}
We derive the low velocity consequences of the Hubble effect in the two gauges we are interested in (Hubble gauge and Riemann gauge) without duplication of efforts. Readers who dislike or even mistrust Weyl geometry can easily follow the argument for the classical case by putting $\varphi = 0$ from the outset, or  know the result anyhow \cite[213ff.]{Weinberg:Cosmology}.
As in the main text, we use  index denotations  $\alpha , \beta , \gamma  = 1,2, 3, \; i, j, k = 0, 1,2,3$. We consider a timelike curve $c(s)$ in proper time parametrization $s $ in a coordinate system  $(x^0, x^1, x^2, x^3)$, with $x^0 =:t$  the coordinate time.  For the inverse function $s(t)$ of $t(s)$ we have, in abbreviated notation, $s' t' = 1$ and thus:
\[ \frac{d^2 x^{\alpha }}{dt^2} =
 \frac{ds}{dt} \frac{d}{ds} (\frac{ds}{dt} \frac{dx^{\alpha }}{ds}) = 
\left( \frac{ds}{dt}\right)^2 \frac{d^2 x^{\alpha }}{ds^2} -
 \left(\frac{dt}{ds}\right)^{-3}\frac{d^2 t}{ds^2}\frac{dx^{\alpha }}{ds} \, 
\]
 If $c$ is a timelike geodesic with  $c'=u$, $|u'|^2= -1$, it  satisfies 
(equ. (\ref{def-geodesic})) and therefore
\beqa  \frac{d^2 x^{\alpha }}{dt^2} &= & 
\left( \frac{ds}{dt}\right)^2  \left(  - \Gamma^{\alpha }_{ij} \frac{dx^i}{ds}\frac{dx^j}{ds} + \varphi_i \frac{dx^{\alpha }}{ds} \frac{dx^i}{ds}  \right) \\
& & - \left( \frac{ds}{dt}\right)^3 \frac{dx^{\alpha}}{ds} \left( -\Gamma^0_{ij} \frac{dx^i}{ds}\frac{dx^j}{ds} + \varphi_i \frac{dt}{ds} \frac{dx^i}{ds}  \right) \\
&=&    - \Gamma^{\alpha }_{ij} \frac{dx^i}{dt}\frac{dx^j}{dt} + \varphi_i \frac{dx^{\alpha }}{dt} \frac{dx^i}{dt} 
+ \Gamma^0_{ij} \frac{dx^i}{dt}\frac{dx^j}{dt} \frac{dx^{\alpha}}{dt} - \varphi_i  \frac{dx^i}{dt} \frac{dx^{\alpha}}{dt} 
\eeqa
 \noindent
We thus get the equation of motion for mass points in eGRT, parametrized  in coordinate time:
\beqarr \label{equ of motion}
\frac{d^2 x^{\alpha }}{dt^2} &= & 
- \Gamma^{\alpha}_{00} +\Gamma^{0}_{00} \frac{dx^{\alpha}}{dt}
 - 2 \Gamma^{\alpha}_{0 \beta} \frac{dx^{\beta}}{dt}
-  \Gamma^{\alpha}_{ \beta \gamma} \frac{dx^{\beta}}{dt}\frac{dx^{\gamma}}{dt}  \\
& & + 2  \Gamma^{0}_{0 \beta } \frac{dx^{\alpha}}{dt}\frac{dx^{\beta}}{dt}
+ \Gamma^{0}_{ \beta \gamma } \frac{dx^{\alpha}}{dt}\frac{dx^{\beta}}{dt}\frac{dx^{\gamma}}{dt}
\nonumber 
\eeqarr

Note that the length connection terms which come from the scale covariance modification of  the geodesic equation cancel. This is expression of the fact that only the trace of the geodesic enters into the coordinate time parametrization of the dynamical equation (\ref{equ of motion}).
In the result  the dynamics of mass points in eGRT is governed by  the guiding field (the affine connection), like in the classical semi-Riemannian case 
 \cite[equ. (9.1.2)]{Weinberg:Cosmology}. Here, of course, the {\em  length connection enters into the affine connection} and influences the dynamics. With it, the choice of a physically ``correct''  (matter) gauge can in principle be read off from the dynamics of mass points. 

\subsection*{Appendix III: Cosmological corrections for   low velocity orbits}
If we want to study the tiny cosmological effects on motions inside the solar system or of other low velocity motions like  clusters dynamcis, we  have to  superimpose  cosmological terms on the weak field approximations used in the respective context. We do so for both,  the classical case and  the Weyl geometric, Hubble gauged version of Robertson-Walker manifolds. To fix the imagination, we start from considering   the relativistic Newton potential $g = diag (-1-2\Phi , 1,1,1)$ with $\Phi (x) = - N \frac{ M}{r}$ and superpose the cosmological constribution on the guiding field (the affine connection). The  argument holds, however, more in general, for all ``small''  relativistic corrections in  post-Newtonian approximations. 

In the  {\em Friedmann-Lema\^{\i}tre} picture,  the Hubble effect arises mathematically from the  warp function in the metric
\[ ds^2 = [c^2]  dt^2 - f^2 (t) d\sigma^2 \, .  \]
Here the spatial component of the metric is usually expressed in spherical coordinates $(x^1,x^2,x^3) = (r, \theta, \varphi)$ (here $\varphi$ denotes, of course, an angle, no connection). The only non-vanishing Christoffel symbols which enter the low velocity approximation are 
\[  \Gamma^{\alpha}_{0, \alpha } =\frac{f'}{f} \, , \]
because, among others,  $\Gamma^{0}_{0 0} = 0$ and $\Gamma^{\alpha}_{0 \beta} = 0$  for $\alpha \not= \beta $. 
Of course the Weylian length connection term in  (\ref{lv equation})  vanishes, but the third term of 
 the low velocity equation gives a cosmological correction to the solar system contribution, with $\frac{f' }{f} = H$.  In  the Newtonian case we get for $t = t_0, H(t_0) = H_0$:
\begin{eqnarray} \label{LowStandard}  \frac{d^2 x^{\alpha }}{dt^2}\approx   - {} _g\Gamma^{\alpha}_{00}  
 - 2 \, {}_g\Gamma^{\alpha}_{0 \beta} \frac{dx^{\beta}}{dt} \nonumber \\
\approx    -    \partial_{\alpha} \Phi - 2\, H_0  \frac{dx^{\alpha }}{dt}  \end{eqnarray}
Thus there is a {\em  low velocity effect of the expandings space cosmology, if it is taken
 seriously.}\footnote{Some astrophysicists seem to believe that inside galaxies the space expansion is somehow suspended and  the warp function has    a realistic interpretation only in intergalactic space. In this (in my view ``unserious'') interpretation of the expanding space cosmology, the Hubble effect could not have any solar system effect at all. 
 \label{serious}} 
It consists of an acceleration proportional, but opposite, to the velocity of the object with regard to a cosmological comoving  coordinate frame with factor $ 2H_0$.  

 If  a post-Newtonian weak field approximation, for example derived from the Schwarzschild solution,    leads to the r.h.s $ \Psi(x, x')$ of the equation, the cosmological correction of a 
Friedmann-Lema\^{\i}tre model is  formally the same
\begin{equation}  \label{lv Friedmann}\frac{d^2 x^{\alpha }}{dt^2}  \approx   \Psi(x, x') - 2 H_0  \frac{dx^{\alpha }}{dt} \, .
  \end{equation}

In the {\em Hubble gauged} Weyl geometric view of {\em Robertson Walker manifolds}, equ. (\ref{Hubble gauge}),  the  affine connection, or more precisely its cosmological contribution,  is 
\begin{eqnarray*} \Gamma^0_{00} = \Gamma^{\alpha}_{\alpha 0} = f' \, ,\;\;\; 
\Gamma ^{0}_{\alpha \alpha } =  g_{\alpha \alpha } f' \, ,\;\;\;
\Gamma ^{\alpha }_{\beta \gamma } = 
\tilde{\Gamma} ^{\alpha }_{\beta \gamma } \, ,
\end{eqnarray*}
with $\tilde{\Gamma} ^{\alpha }_{\beta \gamma }$ the Christoffel symbols of the (constant) space fibre $S_{\kappa }$. All other components of $\Gamma $ vanish. 
The Hubble parameter is $H = \varphi_0 = f'$ and $f'(t_0) = H_0$. The  superposition with the solar system weak field approximation like above yields  the low velocity equation
\begin{equation} \label{lv Hubble gauge} \frac{d^2 x^{\alpha }}{dt^2}  \approx   \Psi(x, x') -  H_0  \frac{dx^{\alpha }}{dt} \, .
\end{equation}
Thus the low velocity acceleration effect is in form comparable to the one in the Friedmann-Lema\^{\i}tre approach. But here it is  due to the Hubble connection  and only half the value of the classical model. In particular, the ``cosmological'' low velocity correction is {\em not dependent on the purely spatial curvature} $\kappa$.
 For {\em Weyl universes} the situation is particularly simple;  here $f'=H$ is a true constant,  the Hubble constant $H_0$.

There is no disagreement in present astrophyscis {\em that} cosmological correction terms have to be taken into account for  the estimation of dynamical  masses  on the level of galaxy clusters.  There are, however, different approaches for the calculation of the {\em how} the correction should be calculated. In  \cite{Peebles:LargeScale} we find a striking heuristic derivation of a cosmological correction term for accelerations of a low velocity motion (equ. (14.1) ff.). Here a description of a  motion $x(t)$  to be expected by dynamical laws in a local coordinate system without cosmic expansion (``peculiar'' motion in Peebles' language),  is compared to a corresponding motion  $\tilde{x}(t) $ expected in an expanding cosmology with scale factor $a(t)$. Because in $\tilde{x}'(t) = a' x + a x'$ the first term may be neglected in approximations, the author arrives at $ \tilde{x}'' = a(  x'' + \frac{a'}{a}) x'$. The dynamically effective acceleration is then
\[ x'' = \frac{1}{a} \tilde{x}''  - H_0 x' \; , \] 
with $\frac{a'}{a} (t_0)= H_0$. 

The result of  this intuitively convincing argument differs  from the  cosmological correction term of the Friedmann-Lema\^{\i}tre context, but  agrees with the low velocity approximation in IWG.
 As \cite{Peebles:LargeScale} may be considered  an authoritative reference work for  theoretical methods used in empirical determination of dynamical mass densities, we conclude that the {\em  actual practice of evaluation of dynamical mass data is  consistent with the framework of IWG, but it is inconsistent with the Friedmann-Lema\^{\i}tre approach}. In the latter  the cosmological correction term for low velocity orbits had to be twice as large.
%%%%%%%%%%%%%%
\vspace{20mm}\\
%\nocite{Weyl:GA}
\newpage
\footnotesize
 \bibliographystyle{apsr}
  \bibliography{a_litfile}

\end{document}